\documentclass[prd,twocolumn,twoside,preprintnumbers,superscriptaddress,nofootinbib]{revtex4-2}
\usepackage{physics}
\usepackage{amsmath}
\usepackage{amssymb}
\usepackage{hyperref}
\usepackage{enumitem}
\usepackage{xspace}
\usepackage{slashed}
\usepackage{braket}
\usepackage{leftindex}
\hypersetup{
    colorlinks=true,       
    linkcolor=blue,          
    citecolor=blue,        
    filecolor=blue,      
    urlcolor=blue           
}
\usepackage{graphicx}  %
\usepackage{bm}  %
\usepackage{amsmath}   %
\usepackage{graphics}
\usepackage{color}
\usepackage{colortbl}
\usepackage{ulem}
\usepackage{hyperref}
\usepackage{xcolor}
\usepackage{comment}
\usepackage{enumitem}

\newcommand{\beq}{\begin{equation}}
\newcommand{\eeq}{\end{equation}}
\newcommand{\bea}{\begin{eqnarray}}
\newcommand{\eea}{\end{eqnarray}}

\newcommand{\dslash}[1]{#1 \llap{/\kern-0.5pt}}
\newcommand{\Dslash}[1]{#1 \llap{/\kern+1.5pt}}
\newcommand{\DDslash}[1]{#1 \llap{/\kern+2.3pt}}
\newcommand{\dslashh}[1]{#1 \llap{/\kern+1pt}}

\newcommand{\be}{\begin{equation}}
\newcommand{\ee}{\end{equation}}
\newcommand{\bma}{\begin{pmatrix}}
\newcommand{\ema}{\end{pmatrix}}

\newcommand{ \mysmall}[1]{\scriptscriptstyle #1} 

\allowdisplaybreaks[1]

\begin{document}

\title{Model Independent Tests of the 
Hadronic Vacuum Polarization \\
Contribution to the Muon \boldmath{$g$$-$$2$}}

\author{Luca~Di~Luzio}
\affiliation{Istituto Nazionale di Fisica Nucleare, Sezione di Padova, Via F.~Marzolo 8, 35131 Padova, Italy}

\author{Alexander~Keshavarzi}
\affiliation{Department of Physics and Astronomy, The University of Manchester, Manchester M13 9PL, U.K.}

\author{Antonio~Masiero}
\affiliation{Dipartimento di Fisica e Astronomia `G.~Galilei', Universit\`a di Padova, Via F.~Marzolo 8, 35131 Padova, Italy}
\affiliation{Istituto Nazionale di Fisica Nucleare, Sezione di Padova, Via F.~Marzolo 8, 35131 Padova, Italy}

\author{Paride~Paradisi}
\affiliation{Dipartimento di Fisica e Astronomia `G.~Galilei', Universit\`a di Padova, Via F.~Marzolo 8, 35131 Padova, Italy}
\affiliation{Istituto Nazionale di Fisica Nucleare, Sezione di Padova, Via F.~Marzolo 8, 35131 Padova, Italy}

\begin{abstract}

The hadronic vacuum polarization (HVP) contributions to the muon $g$$-$$2$ are the crucial quantity to resolve whether new physics is present or not in the comparison  between the Standard Model (SM) prediction and experimental measurements at Fermilab. They are commonly and historically determined via dispersion relations using a vast catalogue of experimentally measured, low-energy $e^+e^-\to \,\rm{hadrons}$ cross section data as input. These dispersive estimates result in a SM prediction that exhibits a muon $g$$-$$2$ discrepancy of more than $5\sigma$ when compared to experiment. However, recent lattice QCD evaluations of the HVP and a new hadronic cross section measurement from the CMD-3 experiment favor a no-new-physics scenario and, therefore, exhibit a common tension with the previous $e^+e^-\to \,\rm{hadrons}$ data. This study explores the current and future implications of these two scenarios on other observables that are also sensitive to the HVP contributions in the hope that they may provide independent tests of the current tensions observed in the muon $g$$-$$2$. 

\end{abstract}

\maketitle

{\it Introduction.---}
\label{sec:intro}
The anomalous magnetic moment of the muon, $a_\mu \equiv (g_\mu -2)/2$, stands amongst the best probes of the Standard Model (SM) of particle physics and its possible extensions. Recently, the Muon $g$$-$$2$ Experiment at Fermilab announced a new experimental (exp) measurement of $a_\mu$~\cite{Muong-2:2023cdq,Muong-2:2024hpx} which, when combined with the previous and consistent results from the same experiment~\cite{Muong-2:2021ojo,Muong-2:2021xzz,Muong-2:2021ovs,Muong-2:2021vma} (and from the earlier Muon $g$$-$$2$ Experiment at Brookhaven~\cite{Muong-2:2006rrc,Muong-2:2002wip,Muong-2:2004fok}), results in a new world average of $a^{\rm exp}_\mu = 116592059(22)\times 10^{-11}$ with an unprecedented precision of 190 parts-per-billion (ppb). 

The most recent, community-approved SM prediction of $a_\mu$ from the Muon $g$$-$$2$ Theory Initiative~\cite{Aoyama:2020ynm,aoyama:2012wk,Aoyama:2019ryr,czarnecki:2002nt,gnendiger:2013pva,davier:2017zfy,keshavarzi:2018mgv,colangelo:2018mtw,hoferichter:2019mqg,davier:2019can,keshavarzi:2019abf,kurz:2014wya,melnikov:2003xd,masjuan:2017tvw,Colangelo:2017fiz,hoferichter:2018kwz,gerardin:2019vio,bijnens:2019ghy,colangelo:2019uex,Blum:2019ugy,colangelo:2014qya} relies on the hadronic vacuum polarization (HVP) contribution, $a_\mu^{\rm \mysmall HVP}$, being entirely evaluated using the data-driven, dispersive approach~\cite{Brodsky:1967sr,Lautrup:1968tdb,Krause:1996rf,kurz:2014wya,Jegerlehner:2017gek,Jegerlehner:2015stw,Jegerlehner:2017lbd,Jegerlehner:2017zsb,Jegerlehner:2018gjd,Eidelman:1995ny,Benayoun:2007cu,Benayoun:2012etq,Benayoun:2012wc,Benayoun:2015gxa,Benayoun:2019zwh,Benayoun:2019zwh,Davier:2010nc,Davier:2010nc,davier:2017zfy,keshavarzi:2018mgv,colangelo:2018mtw,hoferichter:2019mqg,davier:2019can,keshavarzi:2019abf,kurz:2014wya,melnikov:2003xd,masjuan:2017tvw,Colangelo:2017fiz,hoferichter:2018kwz}. This method uses measured $e^+e^- \to \,\rm{hadrons}$ cross section data ($\sigma_{\rm had}$) as input 
and the leading-order (LO) HVP contribution was found by the Theory Initiative to be~\cite{Aoyama:2020ynm,davier:2017zfy,keshavarzi:2018mgv,colangelo:2018mtw,hoferichter:2019mqg,davier:2019can,keshavarzi:2019abf}
\begin{align} 
\label{eq:currentB}
(a_\mu^{\rm \mysmall HVP})_{e^+e^-}^{\rm \mysmall TI} = (693.1 \pm 4.0) \times 10^{-10}\,.
\end{align}
This, when combined with the updated values for the other 
SM contributions, resulted in a SM prediction of 
$a^{\rm\mysmall SM}_\mu = 116591810(43) \times 10^{-11}$~\cite{Aoyama:2020ynm,aoyama:2012wk,Aoyama:2019ryr,czarnecki:2002nt,gnendiger:2013pva,davier:2017zfy,keshavarzi:2018mgv,colangelo:2018mtw,hoferichter:2019mqg,davier:2019can,keshavarzi:2019abf,kurz:2014wya,melnikov:2003xd,masjuan:2017tvw,Colangelo:2017fiz,hoferichter:2018kwz,gerardin:2019vio,bijnens:2019ghy,colangelo:2019uex,Blum:2019ugy,colangelo:2014qya}. Comparing $a^{\rm\mysmall SM}_\mu$ with the current experimental world average leads to
\begin{equation}
\Delta a_\mu \equiv a^{\rm exp}_\mu - a^{\rm\mysmall SM}_\mu = (24.9\pm 4.8)\times 10^{-10} \,, 
\label{eq:5.1sigma}
\end{equation}
yielding a discrepancy of $5.1\, \sigma$ and implying the presence of physics beyond the SM (BSM). 

$a_\mu^{\rm \mysmall HVP}$ can now be calculated from first-principles lattice QCD methods~\cite{Aoyama:2020ynm} 
(see also \cite{Budapest-Marseille-Wuppertal:2017okr,RBC:2018dos,Giusti:2019xct,FermilabLattice:2019ugu,Gerardin:2019rua,chakraborty:2017tqp,blum:2018mom,shintani:2019wai,Aubin:2019usy,giusti:2019hkz,Borsanyi:2020mff,Boccaletti:2024guq}). In particular, the BMW lattice QCD collaboration (BMWc) first computed $a_\mu^{\rm \mysmall HVP}$ with sub per-cent  precision~\cite{Borsanyi:2020mff} and found $(a_\mu^{\rm \mysmall HVP})^{\rm \mysmall BMW}_{\rm \mysmall Lattice} = (707.5 \pm 5.5) \times 10^{-10}$, which is evidently larger than the Theory Initiative value~\cite{Aoyama:2020ynm}. They have since updated their result to~\cite{Boccaletti:2024guq}
\begin{align} 
\label{eq:BMW}
& (a_\mu^{\rm \mysmall HVP})^{\rm \mysmall BMW}_{\rm \mysmall Lattice} = (714.1 \pm 3.3) \times 10^{-10}\,,
\end{align}
which reduces $\Delta a_\mu$ to $0.9\sigma$ (fully consistent with $a^{\rm exp}_\mu$ at the current level of precision) at the expense of generating a $4.0\sigma$ tension with Eq.~\eqref{eq:currentB}.
These results have been partially confirmed by other lattice QCD studies of the so-called window quantities~\cite{RBC:2018dos,Lehner:2020crt,ExtendedTwistedMass:2022jpw,RBC:2023pvn,Kuberski:2024bcj,FermilabLattice:2022izv}, 
which further increase the tension between lattice QCD evaluations 
and $(a_\mu^{\rm \mysmall HVP})_{e^+e^-}^{\rm \mysmall TI}$ in the
intermediate~\cite{Colangelo:2022vok}
as well as long-distance~\cite{RBC:2024fic,Djukanovic:2024cmq} windows.


Increasing the hadronic cross section to accommodate $\Delta a_\mu$ in $(a_\mu^{\rm \mysmall HVP})_{e^+e^-}^{\rm \mysmall TI}$ requires shifts 
below $\sim1$ GeV to avoid inconsistency with the electroweak precision fits~\cite{Passera:2008jk,Crivellin:2020zul,Keshavarzi:2020bfy,Malaescu:2020zuc,Colangelo:2020lcg}. Indeed, a large portion of the differences observed between $(a_\mu^{\rm \mysmall HVP})_{e^+e^-}^{\rm \mysmall TI}$ and lattice QCD evaluations dominantly originate from the light-quark-connected contributions (which in-turn are dominated by $\pi^+\pi^-$)~\cite{Benton:2023dci,Benton:2023fcv}. The potential for light new physics  
affecting the interpretation of $e^+e^-$ data was also investigated in~\cite{DiLuzio:2021uty,Darme:2021huc,Crivellin:2022gfu,Darme:2022yal,Coyle:2023nmi}.

Recently, the CMD-3 experiment released a new measurement of the $e^+e^- \to \pi^+\pi^-$ cross section~\cite{CMD-3:2023alj,CMD-3:2023rfe}. It covers the energy range from 0.32 to 1.2 GeV, therefore capturing the dominant $\rho$-resonance region, and is $\sim2\sigma$-$4\sigma$ higher than all previous $e^+e^- \to \pi^+\pi^-$ cross section measurements in the same region~\cite{Achasov:2006vp,Aulchenko:2006dxz,CMD-2:2006gxt,KLOE:2008fmq,KLOE:2010qei,KLOE:2012anl,KLOE-2:2017fda,BaBar:2009wpw,BaBar:2012bdw,BESIII:2015equ,Xiao:2017dqv,SND:2020nwa}. Although $(a_\mu^{\rm \pi\pi})_{e^+e^-}^{\rm \mysmall CMD3}$ covers a limited energy range, it goes in the same direction as the BMWc result of Eq.~\eqref{eq:BMW}, thereby also suggesting consistency between $a^{\rm exp}_\mu$ and $a^{\rm\mysmall SM}_\mu$. 

\begin{figure*}[t!]
\centering
\includegraphics[width=0.9\linewidth]{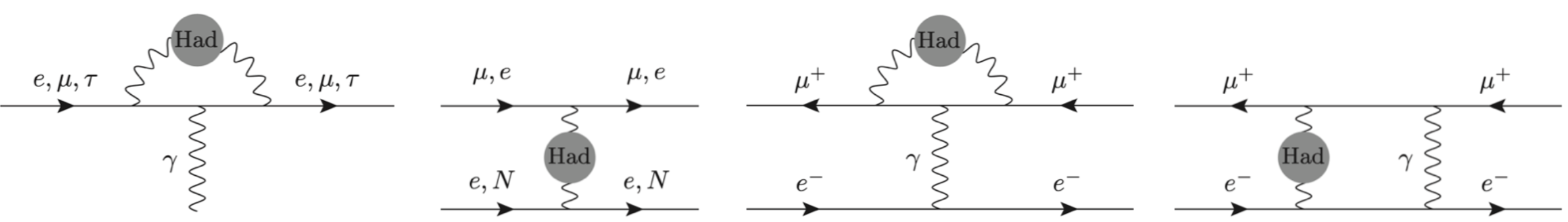}
\caption{
Representative Feynman diagrams for observables 
that are sensitive to the leading HVP contribution: 
leptonic $g$$-$$2$ (first diagram), running of $\alpha$ 
and $\sin^2\theta_W$ (second diagram), and Muonium HFS 
(third and fourth diagrams).}
\label{fig:HVP_Feynman}
\end{figure*}

Several updates and improvements on all fronts are expected soon which may help shed light on the current puzzling picture. New results of the $e^+e^- \to \pi^+\pi^-$ cross section are expected from the BaBar~\cite{newBaBar}, Belle II~\cite{newBelleII}, BESIII~\cite{newBESIII}, CMD-3~\cite{newCMD3}, KLOE~\cite{newKLOE} and SND~\cite{newSND} experiments in the next 2-4 years, with theoretical improvements also expected for the higher-order QED corrections which are paramount for these measurements (see e.g.~\cite{BaBar:2023xiy,Davier:2023fpl}). New full HVP determinations from other lattice QCD collaborations are also expected soon, as well as new dispersive determinations of the HVP incorporating new $e^+e^- \to$ hadrons cross section data. In addition, a direct experimental measurement of $a_\mu^{\rm \mysmall HVP}$, has been proposed and is in preparation by the MUonE experiment~\cite{CarloniCalame:2015obs,Abbiendi:2016xup,Abbiendi:2677471,Banerjee:2020tdt,Masiero:2020vxk}.\footnote{Several efforts have also attempted new determinations of $a_\mu^{\rm \mysmall HVP}$ from hadronic $\tau$-decay data (see e.g.~\cite{Masjuan:2023qsp,Davier:2023fpl}), although the level to which the required isospin breaking corrections are understood is still under question~\cite{Aoyama:2020ynm}.} 

This letter adds to this catalogue of probes other observables that are sensitive to HVP effects and, therefore, can be used as additional indirect tools to scrutinize the differences in $a_\mu^{\rm \mysmall HVP}$. The new data from CMD-3 are used to perform numerical analyses and provide analytical expressions relating the HVP contributions of other observables to $a_\mu^{\rm \mysmall HVP}$. The current and future implications of the CMD-3 data (and, by data-based proxy, the increase of $(a_\mu^{\rm \mysmall HVP})^{\rm \mysmall BMW}_{\rm \mysmall Lattice}$ compared to dispersive estimates) are assessed and quantified for the electron, muon and tau $g$$-$$2$, the running of the QED coupling constant ($\alpha$), the low-energy weak mixing angle ($\sin^2\theta_W(0)$) and the Muonium hyperfine splitting (HFS). The conclusions drawn focus upon how new, precise experimental measurements of these observables could be sensitive to the differences observed due to the CMD-3 data in $a_\mu^{\rm \mysmall HVP}$ and, therefore, provide additional robust tests of the current tensions.

{\it Impact of the CMD-3 data.---}
The CMD-3 measurement of the $\pi^+\pi^-$ final state~\cite{CMD-3:2023alj} is larger than all previous $\pi^+\pi^-$ measurements 
by $2\sigma$-$4\sigma$, resulting in it being comparatively larger than the compilation by the KNT collaboration in 2019 (referred to as KNT19)~\cite{keshavarzi:2019abf}.
Given it is not expected that the differences between these measurements will be reconciled, and that no new compilation of $\pi^+\pi^-$ measurements including the CMD-3 data has yet been performed, two scenarios can be compared and analysed: (1)~$\sigma_{\rm had} = $ KNT19. The previous data are correct and the new CMD-3 data are not used. (2)~$\sigma_{\rm had} = $ KNT19/CMD-3. The CMD-3 data are correct therefore are substituted into and replace the KNT19 data only in their available energy range.

\begin{figure}[t!]
    \centering
   \includegraphics[width=0.47\textwidth]{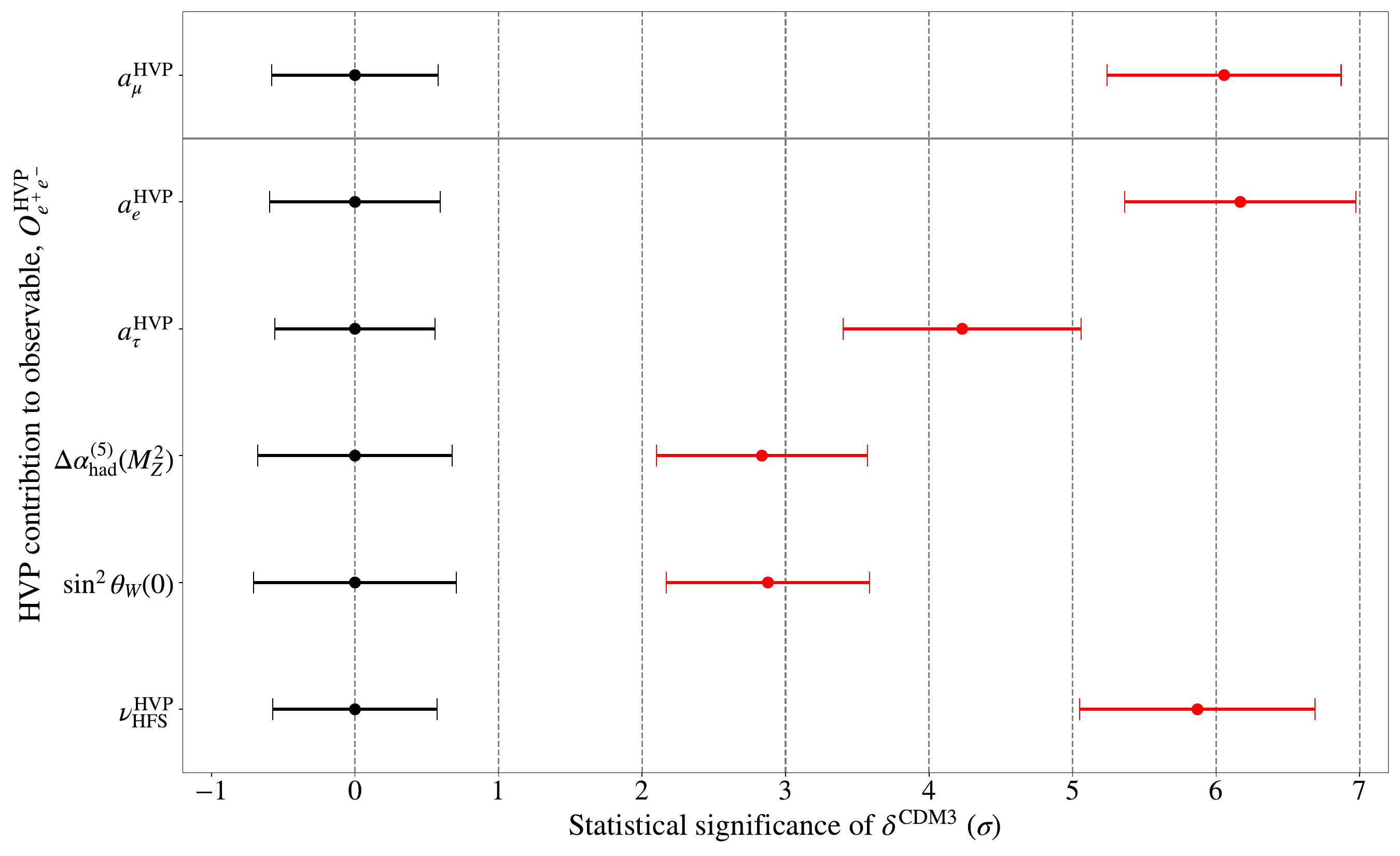}
   \caption{The tension observed in standard deviations ($\sigma$) between dispersive evaluations of HVP contributions to observables sensitive to such effects when the $e^+e^- \rightarrow$ hadrons data used as input as those given in scenario (1) (KNT19)~\cite{keshavarzi:2019abf} (black points) or scenario (2) KNT19/CMD3~\cite{CMD-3:2023alj} (red points). 
   }
   \label{fig:CMD3}
\end{figure}

The electron, muon and tau $g$$-$$2$, the running of the QED coupling constant, the low-energy weak mixing angle and the Muonium HFS are all sensitive to HVP effects (see e.g.~Fig.~\ref{fig:HVP_Feynman}). 
To understand the impact of the CMD-3 data on the HVP contributions to these observables, $O^{\rm \mysmall HVP}_{e^+e^-}$, the difference between the scenario (1) and scenario (2) is calculated as
\begin{align}
\label{eq:shift}
\delta O^{\rm \mysmall CMD3} & \equiv
(O^{\rm \mysmall  HVP}_{e^+e^-})^{\rm \mysmall CMD3} -
(O^{\rm \mysmall  HVP}_{e^+e^-})^{\rm \mysmall KNT19} \, .
\end{align}
All results are given in Table~\ref{table:compareO}. Other than the region of the replaced $\pi^+\pi^-$ data, the two scenarios are correlated for all common KNT19 data. The statistical significance of each $\delta O^{\rm \mysmall CMD3}$ is depicted in Fig.~\ref{fig:CMD3}. The results for each $O^{\rm \mysmall  HVP}_{e^+e^-}$ are discussed in the following, starting with the muon $g$$-$$2$ as the original source of the tensions that motivates this study.

\begin{table}[t!] 		
\centering
\begin{tabular}{| c | c | c | c |}
\hline
$O^{\rm \mysmall  HVP}_{e^+e^-}$ & 
Scenario (1) &
Scenario (2) & 
$\delta O^{\rm \mysmall CMD3}$   
\\
\hline
$a^{\rm\mysmall HVP}_\mu \times 10^{10}$ &
$692.8\,\pm\,2.4$  &
$714.5\,\pm\,3.4$   & 
$21.7\,\pm\,3.6$\hphantom{2}
\\ 
\hline
$a^{\rm\mysmall HVP}_e \times 10^{14}$ &
$186.1\,\pm\,0.7$  &
$192.0\,\pm\,0.9$  & 
$6.0\,\pm\,1.0$ 
\\
\hline
$a^{\rm\mysmall HVP}_\tau \times 10^{8}$ &
$332.8\,\pm\,1.4$  &
$340.2\,\pm\,2.1$   & 
$7.4\,\pm\,1.7$ 
\\ 
\hline
$\Delta\alpha^{(5)}_{\rm had}(M^2_Z) \!\times\! 10^{4}$ &
$276.1\,\pm\,1.1$  &
$277.5\,\pm\,1.2$  & 
$1.4\,\pm\,0.5$ 
\\
\hline
$\sin^2\theta_W(0) \!\times\! 10^4$ &
$2386.0\,\pm\,1.4$\hphantom{2}  &
$2386.4\,\pm\,1.4$\hphantom{2}  & 
$0.4\,\pm\, 0.1$ 
\\ 
\hline
$\nu^{\rm\mysmall HVP}_{\rm\mysmall HFS}$ (Hz) &
$540.5\,\pm\,1.9$  &
$557.0\,\pm\,2.7$  &
$16.5\,\pm\,2.8$\hphantom{1} 
\\
\hline
\end{tabular}
\caption{HVP contributions to the studied observables [first column] using scenario (1) KNT19 data [second column], or scenario (2) KNT19/CMD3 data [third column]. 
The fourth column lists the difference or shift induced by the CMD-3 data as $\delta O^{\rm \mysmall CMD3} \equiv
(O^{\rm \mysmall  HVP}_{e^+e^-})^{\rm \mysmall CMD3} -
(O^{\rm \mysmall  HVP}_{e^+e^-})^{\rm \mysmall KNT19}$, where the error accounts for the correlation between scenario (1) and scenario (2) from the common KNT19 data.
}
\label{table:compareO}
\end{table}

{\it Muon $g$$-$$2$.---}
\label{sec:muongm2}
The LO HVP contribution to the muon $g$$-$$2$ is calculated via
\begin{align} 
\label{eq:sigmatoab}
(a_\mu^{\rm \mysmall HVP})_{e^+e^-} &= \frac{1}{4\pi^3} \int_{m^2_{\pi^0}}^{\infty} 
\text{d}s \, K_\mu(s) \,\sigma_{\rm had}(s) \, . 
\end{align} 
The positive-definite kernel function $K_\mu(s)$ is given by
\begin{equation}
\label{eq:Ks}
K_\mu(s) = \int^{1}_{0} dx
\frac{x^2(1-x)}{x^2 + (1-x)s/m^2_\mu}\,,
\end{equation}
with $K_\mu(s) \approx m^2_\mu / 3s$ in the high-energy limit
$\sqrt{s} \gg m_\mu$. 

The KNT19 evaluation (scenario (1)) was one of several inputs to the estimate found in Eq.~\eqref{eq:currentB} and found the LO HVP contribution to be~\cite{keshavarzi:2019abf}
\begin{align} 
\label{eq:KNT19}
&(a_\mu^{\rm \mysmall HVP})_{e^+e^-}^{\rm \mysmall KNT19} = (692.8 \pm 2.4) \times 10^{-10}
\,.
\end{align}
Using scenario (2) (substituting in the CMD-3 data) \cite{CMD-3:2023alj} increases this estimate to
\begin{align} 
\label{eq:CMD3}
&(a_\mu^{\rm \mysmall HVP})_{e^+e^-}^{\rm \mysmall CMD3} = (714.5 \pm 3.4) \times 10^{-10}\,,
\end{align}
which is in better agreement with the BMWc value of Eq.~(\ref{eq:BMW}). This results in a shift in $a^{\rm\mysmall HVP}_\mu$ of
\begin{align} 
\label{eq:CMD3_shift}
\delta a_\mu^{\rm \mysmall CMD3} &\equiv
(a_\mu^{\rm \mysmall HVP})_{e^+e^-}^{\rm \mysmall CMD3} -
(a_\mu^{\rm \mysmall HVP})_{e^+e^-}^{\rm \mysmall KNT19}
\nonumber\\
&= (21.7 \pm 3.6) \times 10^{-10} \, ,
\end{align}
exhibiting a tension of $6.1\sigma$. In the following, the phenomenological implications of the above shift on 
$O^{\rm \mysmall  HVP}_{e^+e^-}$ (see Table~\ref{table:compareO}) are analyzed, followed by a discussion regarding the required experimental and theoretical improvements needed to probe the disagreement in Eq.~(\ref{eq:CMD3_shift}).

{\it Electron $g$$-$$2$.---}
\label{sec:egm2}
The anomalous magnetic moment of the electron, $a_e$, is commonly used to determine the value of the fine-structure constant, $\alpha$ (and vice versa). Recent, more-refined atomic-physics experiments using Cesium (Cs) and Rubidium (Rb) interferometry have led to the following results for direct measurements of $\alpha$:
\begin{align}
\alpha({\rm Cs}) &= 1/137.035999046(27) \quad \text{\cite{Parker:2018vye}} \, , \\
\alpha({\rm Rb}) &= 1/137.035999206(11) \quad \text{\cite{Morel:2020dww}} \,,
\end{align}
in disagreement with each other by 5.5 $\sigma$. Yet, $a_e$ can be extracted and used as another precision test of the SM and its possible extensions~\cite{Giudice:2012ms,Crivellin:2018qmi}.
%
%
Comparing the resulting SM predictions, $a^{\rm SM}_e$, with the latest experimental measurement of $a^{\rm exp}_e = (115\,965\,218\,059~\pm~13) \times 10^{-14}$~\cite{Fan:2022eto}, yields
\begin{align}
\label{eq:daeCs} 
(\Delta a_e)_{\rm Cs} &= (-102.0 \pm 26.4)\times 10^{-14} \Rightarrow -3.9\sigma \, , \\
\label{eq:daeRb}
(\Delta a_e)_{\rm Rb} &= \,\,\,\,\,\,\,\,(33.8 \pm 16.1)\times 10^{-14} \Rightarrow \,\,\, \,2.1\sigma \,,
\end{align}
where $\Delta a_e \equiv a_e^{\rm exp} - a_e^{\rm SM}$.

The impact of the CMD-3 data on the HVP contribution to $a^{\rm SM}_{e}$ can first be estimated by reconsidering the high-energy approximation of Eq.~\eqref{eq:Ks} and relating $a_\mu^{\rm\mysmall HVP}$ and $a_e^{\rm\mysmall HVP}$ by a quadratic-scaling between 
the lepton masses. In doing so, and by employing the value of $\delta a_\mu^{\rm\mysmall CMD3}$ quoted in Eq.~(\ref{eq:CMD3_shift}), 
results in
%
%
\begin{align}
\label{eq:dae_scaling}
\delta a_e^{\rm\mysmall CMD3} 
\approx \delta a_\mu^{\rm\mysmall CMD3} 
\left( \frac{m_e}{m_\mu} \right)^{\!2} 
\!\approx \, (5.1 \pm 0.8) \times 10^{-14} \,.
\end{align}
This is in good agreement with the numerical result in Table~\ref{table:compareO}, which shows a $6.2\sigma$ shift from the CMD-3 data compared to the previous data. It follows that the electron $g$$-$$2$ will only be sensitive to the increase observed on $a^{\rm\mysmall HVP}_\mu$ if the current discrepancy in the measurements of $\alpha$ is resolved and a precision at the level of $\mathcal{O}(10^{-14})$ is achieved for $\Delta a_e$. 

Consider the sources of error in the determinations of $(\Delta a_e)_{\rm Cs}$ and $(\Delta a_e)_{\rm Rb}$:
\begin{table}[!htb] 		
\begin{tabular}{ c c  c }
\hline
$\Delta a_e$ error source & 
Value &
\, \, \, \% of $\Delta a_e$ error \, \, 
\\
\hline
\, Five-loop QED, $\delta a_e^{\rm QED5}$ \, \, & $6 \times 10^{-14}$ &  $5\%\,({\rm Cs})/13\%\,({\rm Rb})$ \\
Hadronic, $\delta a_e^{\rm HAD}$ & $1 \times 10^{-14}$ &  $<1\%$ \\
$\alpha({\rm Cs})$, $\delta a_{e}^{\alpha({\rm Cs})}$ & $22 \times 10^{-14}$ &  $70\%$ \\
$\alpha({\rm Rb })$, $\delta a_{e}^{\alpha({\rm Rb})}$ & $9 \times 10^{-14}$ &  $28\%$ \\
Experiment, $\delta a_e^{\rm exp}$ & $13 \times 10^{-14}$ &  $24\%\,({\rm Cs})/59\%\,({\rm Rb})$ \\
\hline
\end{tabular}
\label{table:aeErrors}
\vspace{-0.3cm}
\end{table}
\noindent
\vspace{8pt}

It follows that should the experimental resolutions on $\alpha$ and $a^{\rm exp}_e$ improve by roughly one order of magnitude, the total uncertainties of $\Delta a_e$ should reach the $\mathcal{O}(10^{-14})$ level and the study of electron $g$$-$$2$ will become sensitive to the increase observed in $a_\mu^{\rm \mysmall HVP}$ due to CMD-3 (and BMWc). 
Fortunately, an improvement on $\delta a_e^{\rm exp}$ by a factor of $\sim 5$ is expected in the next few years.\footnote{
G.~Gabrielse, private communication.}
On the same timescale, new measurements of $\alpha({\rm Rb})$
and $\alpha({\rm Cs})$ by the Paris and Berkeley collaborations, respectively, will significantly reduce the systematic effects that were dominant in their previous measurements.\footnote{
S.~Guellati-Khélifa and H.~Mueller, private communications.}
Such improvements should be sufficient to probe 
the shift induced by CMD-3 (and BMWc) through the electron $g$$-$$2$.


Additionally, $\delta a_e^{\rm QED5}$ reflects the  current $5\sigma$ discrepancy between the two independent calculations of the gauge-invariant set of 5-loop QED diagrams with no lepton loops by Kinoshita et al.~\cite{Aoyama:2014sxa} and Volkov~\cite{Volkov:2019phy,Volkov:2024yzc}.
As a non-negligible source of uncertainty on $\Delta a_e$, and with expected improvements in the experimental inputs, resolving this difference may soon become crucial in this context.\footnote{S.~Laporta, work in progress.
Preliminary results presented at the 
7th Plenary Workshop of the Muon g-2 Theory Initiative, 09–13 september 2024, KEK, Tsukuba, Japan by M.~Nio 
(see \cite{discrepancyres})
point toward a resolution of this discrepancy.
}

{\it Tau $g$$-$$2$.---}
\label{sec:taugm2}
The shift in $a^{\rm SM}_\tau$ due to the CMD-3 data is estimated 
by assuming dominant effects at the $\rho$-peak:
%
\begin{align} 
\delta a_\tau^{\rm\mysmall CMD3} &\approx 
0.63 \left(m^2_\rho/m^2_\mu \right) \delta a_\mu^{\rm\mysmall CMD3}
\nonumber\\
&\approx (7.2\pm 1.4) \times 10^{-8} \,,
\end{align}
which is in very good agreement with the numerical estimate in Table~\ref{table:compareO}. 
Compared to the electron and muon $g$$-$$2$, the larger mass of the tau lepton increases the weight of the hadronic contributions to higher energies. As such, the influence of the $\pi^+\pi^-$ and $\rho$-resonance contribution is reduced in $(a_\tau^{\rm \mysmall HVP})_{e^+e^-}$, resulting in a much larger degree of correlation between the two scenarios for the tau ($\sim 55\%$) compared to the muon ($\sim 28\%$). This means the significance of $\delta a_\tau^{\rm \mysmall CMD3}$ is $4.2\sigma$ compared to more than $6\sigma$ for $(a_e^{\rm \mysmall HVP})_{e^+e^-}$ and $(a_\mu^{\rm \mysmall HVP})_{e^+e^-}$. 

As $\delta a_\tau^{\rm \mysmall CMD3}$ is roughly one order of magnitude below the expected future experimental sensitivity of $\mathcal{O}(10^{-6})$ at Belle-II~\cite{Bernabeu:2007rr, Bernabeu:2008ii, Chen:2018cxt, Crivellin:2021spu}, an experiment-theory comparison is unlikely to shed light on the current discrepancies in the SM predictions of the muon $g$$-$$2$ without significant improvement in measurements of $a^{\rm exp}_\tau$. 

{\it The running of $\alpha$.---}
\label{sec:alpha}
Hadronic effects to the running QED coupling at the $Z$-boson mass, $\Delta \alpha^{(5)}_{\rm had}(M^2_Z)$, are a principal component of the electroweak precision fit. 
The shift in $\Delta \alpha^{(5)}_{\rm had}(M^2_Z)$ due to the CMD-3 data is estimated as
%
\begin{align}
\delta\Delta \alpha^{(5)}_{\rm had}(M^2_Z) 
&\approx 
\frac{3\pi}{\alpha} \,\frac{m^2_\rho}{m^2_\mu} 
\,\delta a^{\rm\mysmall CMD3}_\mu 
\nonumber\\
&\approx (1.5 \pm 0.3)\times 10^{-4} \,,
\label{eq:shift_alpha}
\end{align}
in full agreement with Table~\ref{table:compareO}. The relevant kernel function in the dispersive evaluation of $\Delta \alpha^{(5)}_{\rm had}(M^2_Z)$ exhibits an $s$-dependent weighting that is much higher in energy than any of the dispersion relations used to evaluate $(a_l^{\rm \mysmall HVP})_{e^+e^-}$. As a result, the CMD-3 increase has less of an impact than in any of the $g$$-$$2$ cases, exhibiting a $2.8\sigma$ deviation. The magnitude of the shift is comparable with the current uncertainty of $\Delta \alpha^{(5)}_{\rm had}(M^2_Z)$, making it difficult for $\Delta \alpha^{(5)}_{\rm had}(M^2_Z)$ to probe $\delta a^{\rm\mysmall CMD3}_\mu $. However, looking ahead, 
a future $e^+e^-$ collider, e.g.~FCC-ee~\cite{Janot:2015gjr,Blondel:2021ema}, is expected to reach an unprecedented precision on $\alpha(M^2_Z)$ of $\mathcal{O}(10^{-5})$, which would provide sensitivity to the shift in Eq.~\eqref{eq:shift_alpha}. 

{\it The running of $\sin^2\theta_W$.---}
\label{sec:sin2W}
The low-energy weak mixing angle, defined as $\sin^2\theta_W(0)$ at zero momentum transfer ($q^2 =0)$, has been measured at parity-violating electron scattering and atomic physics asymmetry experiments~\cite{ParticleDataGroup:2018ovx,Gambino:1993dd}. 
%
Following the prescription in~\cite{Keshavarzi:2020bfy}, $\sin^2\theta_W(0)$ can be connected with the high-energy LEP measurement of the same quantity, $\sin^2\theta_W(M_Z)$, by including quantum effects from the $\gamma-Z$ mixing~\cite{Erler:2017knj,Erler:2024lds}. 

The shift from the CMD-3 data is estimated via\footnote{A similar result was obtained in Ref.~\cite{Erler:2024lds} employing instead 
BMW Lattice data.}
\begin{align}
\label{eq:thetaW_est}
\delta\sin^2\!\theta_W(0) &\approx k' \sin^2\!\theta_W(M_Z)\,
\frac{3\pi}{\alpha} \,\frac{m^2_\rho}{m^2_\mu} \,
\delta a^{\rm\mysmall CMD3}_\mu
\nonumber\\
& 
\approx (0.4 \pm 0.1) \times 10^{-4}
\,,
\end{align}
where $\sin^2\theta_W(M_Z) = 0.23119(14)$~\cite{ParticleDataGroup:2018ovx,Gambino:1993dd} and $k' = 1.14$~\cite{Erler:2017knj,Erler:2024lds}. This is again in full agreement with the numerical estimate in Table~\ref{table:compareO}.
The comparison exhibits a similar significance ($2.9\sigma$) as for $\delta\Delta \alpha^{(5)}_{\rm had}(M^2_Z)$. 

Unfortunately, current and expected future experimental errors prevent $\sin^2\!\theta_W(0)$ to probe the HVP contributions, which would require a precision on both $\sin^2\theta_W(0)$ and $\sin^2\theta_W(M_Z)$ at the~$\sim 10^{-5}$ level. 
Indeed, even a potential $e^+e^-$ high-energy collider, like FCC-ee, would take at least few decades to reach a resolution on $\sin^2\theta_W(M_Z)$ of order $\mathcal{O}(10^{-5})$~\cite{Blondel:2021ema}, a timescale by 
which the puzzle discussed here is very likely to be already successfully overcome. Furthermore, experimental progress on $\sin^2\theta_W(0)$ planned at both MESA (P2) in Mainz~\cite{Becker:2018ggl} and JLab (Møller)~\cite{MOLLER:2014iki} is projected to reach only $\sim\mathcal{O}(10^{-4})$. 
%

{\it The Muonium HFS.---}
\label{sec:Muonium}
\label{sec:hypMuon}
We show now that a powerful method to extract $a^{\rm\mysmall HVP}_\mu$ is via Muonium, 
a bound state of an antimuon and an electron ($M\equiv \mu^+e^-$). 
The Muonium HFS of the 1S ground state is given by~\cite{Tiesinga:2021myr,Eides:2018rph,Czarnecki:2001yx,Karshenboim:2001yy}
%
\beq
\label{eq:MHFSth}
\frac{\nu_{\rm HFS}}{\nu_F} = 1 + 
a_\mu + \Delta^{\rm QED}_{\rm HFS} +
\Delta^{\rm weak}_{\rm HFS} +
\Delta_{\rm HFS}^{\rm HVP}
\,,
\eeq
where 
\begin{equation}
\label{eq:nu_F}
\nu_F = \frac{16}{3} 
\frac{m_e}{m_\mu}
\frac{R_\infty h\, c \, \alpha^2}{(1+m_e/m_\mu)^3}\,,   
\end{equation}
and $R_\infty$ is the Rydberg constant. The quantities $\Delta^{\rm QED}_{\rm HFS}$ and $\Delta^{\rm weak}_{\rm HFS}$ 
in Eq.~(\ref{eq:MHFSth}) refer to corrections stemming from QED and weak contributions, respectively. $\Delta_{\rm HFS}^{\rm HVP}$ accounts for HVP effects~\cite{Czarnecki:2001yx} which are given by~\cite{Czarnecki:2001yx} 
\begin{equation}
\label{eq:muonium_dispersive}
\Delta_{\rm HFS}^{\rm HVP} =
\frac{1}{2\pi^3} 
\int^\infty_{m^2_\pi} ds \, K_{\rm Mu}(s) \,\sigma_{\rm had}(s) \, , 
\end{equation}
where, for $s \gg m^2_\mu$, the kernel function $K_{\rm Mu}(s)$ reads
\begin{equation}
K_{\rm Mu}(s) \approx 
\frac{m^2_\mu}{s}
\left( \frac{9}{2} \log\frac{s}{m^2_\mu} + \frac{15}{4} \right)
\frac{m_e}{m_\mu} 
\,. 
\label{eq:Kmu}
\end{equation}
Again, as the dominant effects to $\sigma_{\rm had}$ arise at the $\rho$-peak, $s \approx m^2_\rho \gg m^2_\mu$. Using Eqs.~\eqref{eq:sigmatoab} and~\eqref{eq:muonium_dispersive} results in
\begin{align}
\label{eq:deltahHVP}
\Delta_{\rm HFS}^{\rm HVP} 
& \approx 
6 \,\frac{m^2_\rho}{m^2_\mu} K_{\rm Mu}(m^2_\rho)\,
(a_{\mu}^{\rm\mysmall HVP})_{e^+e^-} 
\nonumber\\
& \approx 0.63 \, (a_{\mu}^{\rm\mysmall HVP})_{e^+e^-}\,.
\end{align}
Therefore, using also Eq.~(\ref{eq:MHFSth}), the entire HVP contribution to $\nu_{\rm HFS}$ can be expressed as\footnote{
As proposed by Ref.~\cite{Delaunay:2021uph}, Eq.~\eqref{eq:MHFSth} allows for an independent determination of $a_\mu$ if $e^+e^-$ data are used to determine $\Delta^{\rm\mysmall HVP}_{\rm\mysmall HFS}$.
In this case, the obtained $a_\mu$ cannot be exploited to extract $a_\mu^{\rm\mysmall HVP}$.
Instead, since our aim is to determine $a^{\rm\mysmall HVP}_\mu$ from the muonium HFS, we have to take into account both contributions from the $a_\mu$ and $\Delta_{\rm\mysmall HFS}^{\rm\mysmall HVP}$ terms in Eq.~\eqref{eq:MHFSth}.
}
\begin{equation}
\nu^{\rm HVP}_{\rm HFS} = 
\left( a^{\rm\mysmall HVP}_\mu + \Delta^{\rm HVP}_{\rm HFS}\right) \nu_F
\,\approx\, 
1.63 \, \nu_F \, a_{\mu}^{\rm\mysmall HVP}
\,,
\label{eq:HVP_Muonium}
\end{equation}
where the two terms are depicted in the third and fourth diagrams of Fig.~\ref{fig:HVP_Feynman}. 
Therefore, the impact of the CMD-3 data on $\nu^{\rm\mysmall  HVP}_{\rm\mysmall HFS}$ is estimated to be
\begin{equation}
\delta\nu^{\rm \mysmall CMD3}_{\rm HFS} \approx 
(15.8 \pm 2.6)~{\rm Hz} \,,
\label{eq:HVP_Muonium_CMD3_est}
\end{equation}
in excellent agreement with the numerical evaluation in Table~\ref{table:compareO} and exhibits a $5.9\sigma$ discrepancy in the comparison.
This makes the Muonium HFS among the most sensitive probes 
of $\delta a_\mu^{\rm \mysmall CMD3}$.
%
To be sensitive to this shift would require a measurement at a precision of $\mathcal{O}(1)$ Hz. Fortunately, the MuSEUM experiment at J-PARC aims to reduce the uncertainty of the current measurement $\nu^{\rm exp}_{\rm HFS} = (4 \ 463 \ 302 \ 776 \pm 51) \ {\rm Hz}$~\cite{Mariam:1982bq,Liu:1999iz} by one order of magnitude~\cite{Strasser:2016smg}, well below the shift
of Eq.~\eqref{eq:HVP_Muonium_CMD3_est}.

Other sources of uncertainty are also set to improve. The uncertainty on $\nu_F$ is fully dominated by the $m_e/m_\mu$ ratio (the uncertainties of $R_\infty$ and $\alpha$ are negligible)~\cite{Tiesinga:2021myr}. Here, $m_e/m_\mu$ is obtained from the measurement of the 1S-2S transition frequency in Muonium~\cite{Delaunay:2021uph}, given by
\beq
    \label{eq:M1S2Sth}
    \nu_{\rm 1S-2S} 
=   \frac{3}{4}\frac{R_\infty hc}{(1+m_e/m_\mu)}
    \left[1+\delta_{\rm 1S-2S}\right]\,,
\eeq
where the muon mass enters as a recoil contribution. As the hadronic corrections to $\delta_{\rm 1S-2S}$~\cite{Friar:1998wu} do not spoil the extraction of $m_e/m_\mu$ at the required precision, the dominant uncertainty on $m_e/m_\mu$ arises from the experimental error 
in $\nu_{\rm 1S-2S}$~\cite{Meyer:1999cx}. This induces an error 
on $\nu_F$ of $4\times 10^3$ Hz. Experimental progress is ongoing at the Mu-MASS experiment to improve 
this precision by three orders of magnitude~\cite{Crivelli:2018vfe}. Additionally, the dominant theory uncertainty in $\nu_{\rm HFS}$ arises from unknown three-loop QED contributions~\cite{Eides:2018rph} to $\delta^{\rm QED}_{\rm HFS}$ ($\delta_{\rm 1S-2S}$) and amounts to about 70$\,$Hz (7$\,$Hz)~\cite{Eides:2018rph}. A complete three-loop QED calculation
would bring the related uncertainty to a negligible level~\cite{Eides:2016qhf}.
%

{\it Conclusions and outlook.---}
Recent lattice QCD calculations and new measurements by the CMD-3 collaboration show a significant tension with the low-energy 
$e^+e^-\to \, \text{hadrons}$ data in the determination of the 
leading HVP contribution to the muon $g$$-$$2$. 
Several model-independent tests of the HVP contribution are possible through the electron and tau $g$$-$$2$, the running of $\alpha$ and of the weak mixing angle, and the Muonium HFS. Both numerical analyses based 
on state-of-the-art $e^+e^-\to \, \text{hadrons}$ data and approximate analytical expressions relating the HVP contributions of the considered observables have revealed the electron $g$$-$$2$ and the Muonium HFS to be the most promising prospects (see Fig.~\ref{fig:CMD3}), offering unique opportunities to shed light on the current $g$$-$$2$ puzzle. 

To achieve this ambitious goal, both experimental and theoretical improvements are needed. For the $g$$-$$2$ of the electron, the required precision level of $\mathcal{O}(10^{-14})$ is the expected future resolution for $a_e^{\rm exp}$ and $\alpha$ from atomic physics. 
%
In the case of the Muonium HFS, the MuSEUM experiment is expected to improve the HFS ground-state measurement by one order of magnitude, while the Mu-MASS experiment is expected to determine the $m_e/m_\mu$ ratio with a precision three orders of magnitude better than the present situation. Finally, a better theoretical control of QED effects (five loop contributions in the case of the electron $g$$-$$2$ and three loops in the case of Muonium HFS) is necessary. Although challenging, the above program seems to be feasible and it should be pursued with high priority to resolve significant tensions in one of the most stringent probes of the SM.

\vspace{10pt}

{\it Acknowledgments.---}
We thank 
G.~Colangelo, 
P.~Crivelli, 
R.~Frezzotti,
G.~Gabrielse, 
S.~Guellati-Khélifa,
M.~Incagli, 
K.~Kirch, 
S.~Laporta, 
L.~Lellouch, 
V.~Lubicz, 
H.~Mueller,
M.~Neubert, 
M. Passera,
F.~Piccinini, 
M.~Pospelov, 
N.~Tantalo
and 
G.~Venanzoni
for private communications and useful discussions. 
Special thanks are extended to D.~Nomura and T.~Teubner for their collaboration with AK in producing the KNT19 data. 
The work of AK is supported by The Royal Society (URF$\backslash$R1$\backslash$231503). The work of LDL and PP is supported
by the European Union -- Next Generation EU and
by the Italian Ministry of University and Research (MUR) 
via the PRIN 2022 project n.~2022K4B58X -- AxionOrigins.
This work received funding by the INFN Iniziative Specifiche APINE and TASP and from the European Union's Horizon 2020 research and innovation programme under the Marie Sk\l{}odowska-Curie grant agreements n. 860881 -- HIDDeN, n.~101086085 -- ASYMMETRY. This work was also partially supported by the Italian MUR Departments of Excellence grant 2023-2027 ``Quantum Frontiers". 

\bibliographystyle{apsrev4-2.bst}
\bibliography{PRL_HVP_g-2_v2.bib}

\begin{thebibliography}{138}%
\makeatletter
\providecommand \@ifxundefined [1]{%
 \@ifx{#1\undefined}
}%
\providecommand \@ifnum [1]{%
 \ifnum #1\expandafter \@firstoftwo
 \else \expandafter \@secondoftwo
 \fi
}%
\providecommand \@ifx [1]{%
 \ifx #1\expandafter \@firstoftwo
 \else \expandafter \@secondoftwo
 \fi
}%
\providecommand \natexlab [1]{#1}%
\providecommand \enquote  [1]{``#1''}%
\providecommand \bibnamefont  [1]{#1}%
\providecommand \bibfnamefont [1]{#1}%
\providecommand \citenamefont [1]{#1}%
\providecommand \href@noop [0]{\@secondoftwo}%
\providecommand \href [0]{\begingroup \@sanitize@url \@href}%
\providecommand \@href[1]{\@@startlink{#1}\@@href}%
\providecommand \@@href[1]{\endgroup#1\@@endlink}%
\providecommand \@sanitize@url [0]{\catcode `\\12\catcode `\$12\catcode
  `\&12\catcode `\#12\catcode `\^12\catcode `\_12\catcode `\%12\relax}%
\providecommand \@@startlink[1]{}%
\providecommand \@@endlink[0]{}%
\providecommand \url  [0]{\begingroup\@sanitize@url \@url }%
\providecommand \@url [1]{\endgroup\@href {#1}{\urlprefix }}%
\providecommand \urlprefix  [0]{URL }%
\providecommand \Eprint [0]{\href }%
\providecommand \doibase [0]{https://doi.org/}%
\providecommand \selectlanguage [0]{\@gobble}%
\providecommand \bibinfo  [0]{\@secondoftwo}%
\providecommand \bibfield  [0]{\@secondoftwo}%
\providecommand \translation [1]{[#1]}%
\providecommand \BibitemOpen [0]{}%
\providecommand \bibitemStop [0]{}%
\providecommand \bibitemNoStop [0]{.\EOS\space}%
\providecommand \EOS [0]{\spacefactor3000\relax}%
\providecommand \BibitemShut  [1]{\csname bibitem#1\endcsname}%
\let\auto@bib@innerbib\@empty
\bibitem [{\citenamefont {Aguillard}\ \emph {et~al.}(2023)\citenamefont
  {Aguillard} \emph {et~al.}}]{Muong-2:2023cdq}%
  \BibitemOpen
  \bibfield  {author} {\bibinfo {author} {\bibfnamefont {D.~P.}\ \bibnamefont
  {Aguillard}} \emph {et~al.} (\bibinfo {collaboration} {Muon g-2}),\ }\href
  {https://doi.org/10.1103/PhysRevLett.131.161802} {\bibfield  {journal}
  {\bibinfo  {journal} {Phys. Rev. Lett.}\ }\textbf {\bibinfo {volume} {131}},\
  \bibinfo {pages} {161802} (\bibinfo {year} {2023})},\ \Eprint
  {https://arxiv.org/abs/2308.06230} {arXiv:2308.06230 [hep-ex]} \BibitemShut
  {NoStop}%
\bibitem [{\citenamefont {Aguillard}\ \emph {et~al.}(2024)\citenamefont
  {Aguillard} \emph {et~al.}}]{Muong-2:2024hpx}%
  \BibitemOpen
  \bibfield  {author} {\bibinfo {author} {\bibfnamefont {D.~P.}\ \bibnamefont
  {Aguillard}} \emph {et~al.} (\bibinfo {collaboration} {Muon g-2}),\
  }\href@noop {} {\  (\bibinfo {year} {2024})},\ \Eprint
  {https://arxiv.org/abs/2402.15410} {arXiv:2402.15410 [hep-ex]} \BibitemShut
  {NoStop}%
\bibitem [{\citenamefont {Abi}\ \emph {et~al.}(2021)\citenamefont {Abi} \emph
  {et~al.}}]{Muong-2:2021ojo}%
  \BibitemOpen
  \bibfield  {author} {\bibinfo {author} {\bibfnamefont {B.}~\bibnamefont
  {Abi}} \emph {et~al.} (\bibinfo {collaboration} {Muon g-2}),\ }\href
  {https://doi.org/10.1103/PhysRevLett.126.141801} {\bibfield  {journal}
  {\bibinfo  {journal} {Phys. Rev. Lett.}\ }\textbf {\bibinfo {volume} {126}},\
  \bibinfo {pages} {141801} (\bibinfo {year} {2021})},\ \Eprint
  {https://arxiv.org/abs/2104.03281} {arXiv:2104.03281 [hep-ex]} \BibitemShut
  {NoStop}%
\bibitem [{\citenamefont {Albahri}\ \emph
  {et~al.}(2021{\natexlab{a}})\citenamefont {Albahri} \emph
  {et~al.}}]{Muong-2:2021xzz}%
  \BibitemOpen
  \bibfield  {author} {\bibinfo {author} {\bibfnamefont {T.}~\bibnamefont
  {Albahri}} \emph {et~al.} (\bibinfo {collaboration} {Muon g-2}),\ }\href
  {https://doi.org/10.1103/PhysRevAccelBeams.24.044002} {\bibfield  {journal}
  {\bibinfo  {journal} {Phys. Rev. Accel. Beams}\ }\textbf {\bibinfo {volume}
  {24}},\ \bibinfo {pages} {044002} (\bibinfo {year} {2021}{\natexlab{a}})},\
  \Eprint {https://arxiv.org/abs/2104.03240} {arXiv:2104.03240
  [physics.acc-ph]} \BibitemShut {NoStop}%
\bibitem [{\citenamefont {Albahri}\ \emph
  {et~al.}(2021{\natexlab{b}})\citenamefont {Albahri} \emph
  {et~al.}}]{Muong-2:2021ovs}%
  \BibitemOpen
  \bibfield  {author} {\bibinfo {author} {\bibfnamefont {T.}~\bibnamefont
  {Albahri}} \emph {et~al.} (\bibinfo {collaboration} {Muon g-2}),\ }\href
  {https://doi.org/10.1103/PhysRevA.103.042208} {\bibfield  {journal} {\bibinfo
   {journal} {Phys. Rev. A}\ }\textbf {\bibinfo {volume} {103}},\ \bibinfo
  {pages} {042208} (\bibinfo {year} {2021}{\natexlab{b}})},\ \Eprint
  {https://arxiv.org/abs/2104.03201} {arXiv:2104.03201 [hep-ex]} \BibitemShut
  {NoStop}%
\bibitem [{\citenamefont {Albahri}\ \emph
  {et~al.}(2021{\natexlab{c}})\citenamefont {Albahri} \emph
  {et~al.}}]{Muong-2:2021vma}%
  \BibitemOpen
  \bibfield  {author} {\bibinfo {author} {\bibfnamefont {T.}~\bibnamefont
  {Albahri}} \emph {et~al.} (\bibinfo {collaboration} {Muon g-2}),\ }\href
  {https://doi.org/10.1103/PhysRevD.103.072002} {\bibfield  {journal} {\bibinfo
   {journal} {Phys. Rev. D}\ }\textbf {\bibinfo {volume} {103}},\ \bibinfo
  {pages} {072002} (\bibinfo {year} {2021}{\natexlab{c}})},\ \Eprint
  {https://arxiv.org/abs/2104.03247} {arXiv:2104.03247 [hep-ex]} \BibitemShut
  {NoStop}%
\bibitem [{\citenamefont {Bennett}\ \emph {et~al.}(2006)\citenamefont {Bennett}
  \emph {et~al.}}]{Muong-2:2006rrc}%
  \BibitemOpen
  \bibfield  {author} {\bibinfo {author} {\bibfnamefont {G.~W.}\ \bibnamefont
  {Bennett}} \emph {et~al.} (\bibinfo {collaboration} {Muon g-2}),\ }\href
  {https://doi.org/10.1103/PhysRevD.73.072003} {\bibfield  {journal} {\bibinfo
  {journal} {Phys. Rev. D}\ }\textbf {\bibinfo {volume} {73}},\ \bibinfo
  {pages} {072003} (\bibinfo {year} {2006})},\ \Eprint
  {https://arxiv.org/abs/hep-ex/0602035} {arXiv:hep-ex/0602035} \BibitemShut
  {NoStop}%
\bibitem [{\citenamefont {Bennett}\ \emph {et~al.}(2002)\citenamefont {Bennett}
  \emph {et~al.}}]{Muong-2:2002wip}%
  \BibitemOpen
  \bibfield  {author} {\bibinfo {author} {\bibfnamefont {G.~W.}\ \bibnamefont
  {Bennett}} \emph {et~al.} (\bibinfo {collaboration} {Muon g-2}),\ }\href
  {https://doi.org/10.1103/PhysRevLett.89.101804} {\bibfield  {journal}
  {\bibinfo  {journal} {Phys. Rev. Lett.}\ }\textbf {\bibinfo {volume} {89}},\
  \bibinfo {pages} {101804} (\bibinfo {year} {2002})},\ \bibinfo {note}
  {[Erratum: Phys.Rev.Lett. 89, 129903 (2002)]},\ \Eprint
  {https://arxiv.org/abs/hep-ex/0208001} {arXiv:hep-ex/0208001} \BibitemShut
  {NoStop}%
\bibitem [{\citenamefont {Bennett}\ \emph {et~al.}(2004)\citenamefont {Bennett}
  \emph {et~al.}}]{Muong-2:2004fok}%
  \BibitemOpen
  \bibfield  {author} {\bibinfo {author} {\bibfnamefont {G.~W.}\ \bibnamefont
  {Bennett}} \emph {et~al.} (\bibinfo {collaboration} {Muon g-2}),\ }\href
  {https://doi.org/10.1103/PhysRevLett.92.161802} {\bibfield  {journal}
  {\bibinfo  {journal} {Phys. Rev. Lett.}\ }\textbf {\bibinfo {volume} {92}},\
  \bibinfo {pages} {161802} (\bibinfo {year} {2004})},\ \Eprint
  {https://arxiv.org/abs/hep-ex/0401008} {arXiv:hep-ex/0401008} \BibitemShut
  {NoStop}%
\bibitem [{\citenamefont {Aoyama}\ \emph {et~al.}(2020)\citenamefont {Aoyama}
  \emph {et~al.}}]{Aoyama:2020ynm}%
  \BibitemOpen
  \bibfield  {author} {\bibinfo {author} {\bibfnamefont {T.}~\bibnamefont
  {Aoyama}} \emph {et~al.},\ }\href
  {https://doi.org/10.1016/j.physrep.2020.07.006} {\bibfield  {journal}
  {\bibinfo  {journal} {Phys. Rept.}\ }\textbf {\bibinfo {volume} {887}},\
  \bibinfo {pages} {1} (\bibinfo {year} {2020})},\ \Eprint
  {https://arxiv.org/abs/2006.04822} {arXiv:2006.04822 [hep-ph]} \BibitemShut
  {NoStop}%
\bibitem [{\citenamefont {Aoyama}\ \emph {et~al.}(2012)\citenamefont {Aoyama},
  \citenamefont {Hayakawa}, \citenamefont {Kinoshita},\ and\ \citenamefont
  {Nio}}]{aoyama:2012wk}%
  \BibitemOpen
  \bibfield  {author} {\bibinfo {author} {\bibfnamefont {T.}~\bibnamefont
  {Aoyama}}, \bibinfo {author} {\bibfnamefont {M.}~\bibnamefont {Hayakawa}},
  \bibinfo {author} {\bibfnamefont {T.}~\bibnamefont {Kinoshita}},\ and\
  \bibinfo {author} {\bibfnamefont {M.}~\bibnamefont {Nio}},\ }\href
  {https://doi.org/10.1103/PhysRevLett.109.111808} {\bibfield  {journal}
  {\bibinfo  {journal} {Phys. Rev. Lett.}\ }\textbf {\bibinfo {volume} {109}},\
  \bibinfo {pages} {111808} (\bibinfo {year} {2012})},\ \Eprint
  {https://arxiv.org/abs/1205.5370} {arXiv:1205.5370 [hep-ph]} \BibitemShut
  {NoStop}%
\bibitem [{\citenamefont {Aoyama}\ \emph {et~al.}(2019)\citenamefont {Aoyama},
  \citenamefont {Kinoshita},\ and\ \citenamefont {Nio}}]{Aoyama:2019ryr}%
  \BibitemOpen
  \bibfield  {author} {\bibinfo {author} {\bibfnamefont {T.}~\bibnamefont
  {Aoyama}}, \bibinfo {author} {\bibfnamefont {T.}~\bibnamefont {Kinoshita}},\
  and\ \bibinfo {author} {\bibfnamefont {M.}~\bibnamefont {Nio}},\ }\href
  {https://doi.org/10.3390/atoms7010028} {\bibfield  {journal} {\bibinfo
  {journal} {Atoms}\ }\textbf {\bibinfo {volume} {7}},\ \bibinfo {pages} {28}
  (\bibinfo {year} {2019})}\BibitemShut {NoStop}%
\bibitem [{\citenamefont {Czarnecki}\ \emph {et~al.}(2003)\citenamefont
  {Czarnecki}, \citenamefont {Marciano},\ and\ \citenamefont
  {Vainshtein}}]{czarnecki:2002nt}%
  \BibitemOpen
  \bibfield  {author} {\bibinfo {author} {\bibfnamefont {A.}~\bibnamefont
  {Czarnecki}}, \bibinfo {author} {\bibfnamefont {W.~J.}\ \bibnamefont
  {Marciano}},\ and\ \bibinfo {author} {\bibfnamefont {A.}~\bibnamefont
  {Vainshtein}},\ }\href {https://doi.org/10.1103/PhysRevD.67.073006}
  {\bibfield  {journal} {\bibinfo  {journal} {Phys. Rev.}\ }\textbf {\bibinfo
  {volume} {D67}},\ \bibinfo {pages} {073006} (\bibinfo {year} {2003})},\
  \bibinfo {note} {[Erratum: Phys. Rev. {\bf D73}, 119901 (2006)]},\ \Eprint
  {https://arxiv.org/abs/hep-ph/0212229} {arXiv:hep-ph/0212229 [hep-ph]}
  \BibitemShut {NoStop}%
\bibitem [{\citenamefont {Gnendiger}\ \emph {et~al.}(2013)\citenamefont
  {Gnendiger}, \citenamefont {St{\"o}ckinger},\ and\ \citenamefont
  {St{\"o}ckinger-Kim}}]{gnendiger:2013pva}%
  \BibitemOpen
  \bibfield  {author} {\bibinfo {author} {\bibfnamefont {C.}~\bibnamefont
  {Gnendiger}}, \bibinfo {author} {\bibfnamefont {D.}~\bibnamefont
  {St{\"o}ckinger}},\ and\ \bibinfo {author} {\bibfnamefont {H.}~\bibnamefont
  {St{\"o}ckinger-Kim}},\ }\href {https://doi.org/10.1103/PhysRevD.88.053005}
  {\bibfield  {journal} {\bibinfo  {journal} {Phys. Rev.}\ }\textbf {\bibinfo
  {volume} {D88}},\ \bibinfo {pages} {053005} (\bibinfo {year} {2013})},\
  \Eprint {https://arxiv.org/abs/1306.5546} {arXiv:1306.5546 [hep-ph]}
  \BibitemShut {NoStop}%
\bibitem [{\citenamefont {Davier}\ \emph {et~al.}(2017)\citenamefont {Davier},
  \citenamefont {Hoecker}, \citenamefont {Malaescu},\ and\ \citenamefont
  {Zhang}}]{davier:2017zfy}%
  \BibitemOpen
  \bibfield  {author} {\bibinfo {author} {\bibfnamefont {M.}~\bibnamefont
  {Davier}}, \bibinfo {author} {\bibfnamefont {A.}~\bibnamefont {Hoecker}},
  \bibinfo {author} {\bibfnamefont {B.}~\bibnamefont {Malaescu}},\ and\
  \bibinfo {author} {\bibfnamefont {Z.}~\bibnamefont {Zhang}},\ }\href
  {https://doi.org/10.1140/epjc/s10052-017-5161-6} {\bibfield  {journal}
  {\bibinfo  {journal} {Eur. Phys. J.}\ }\textbf {\bibinfo {volume} {C77}},\
  \bibinfo {pages} {827} (\bibinfo {year} {2017})},\ \Eprint
  {https://arxiv.org/abs/1706.09436} {arXiv:1706.09436 [hep-ph]} \BibitemShut
  {NoStop}%
\bibitem [{\citenamefont {Keshavarzi}\ \emph {et~al.}(2018)\citenamefont
  {Keshavarzi}, \citenamefont {Nomura},\ and\ \citenamefont
  {Teubner}}]{keshavarzi:2018mgv}%
  \BibitemOpen
  \bibfield  {author} {\bibinfo {author} {\bibfnamefont {A.}~\bibnamefont
  {Keshavarzi}}, \bibinfo {author} {\bibfnamefont {D.}~\bibnamefont {Nomura}},\
  and\ \bibinfo {author} {\bibfnamefont {T.}~\bibnamefont {Teubner}},\ }\href
  {https://doi.org/10.1103/PhysRevD.97.114025} {\bibfield  {journal} {\bibinfo
  {journal} {Phys. Rev.}\ }\textbf {\bibinfo {volume} {D97}},\ \bibinfo {pages}
  {114025} (\bibinfo {year} {2018})},\ \Eprint
  {https://arxiv.org/abs/1802.02995} {arXiv:1802.02995 [hep-ph]} \BibitemShut
  {NoStop}%
\bibitem [{\citenamefont {Colangelo}\ \emph {et~al.}(2019)\citenamefont
  {Colangelo}, \citenamefont {Hoferichter},\ and\ \citenamefont
  {Stoffer}}]{colangelo:2018mtw}%
  \BibitemOpen
  \bibfield  {author} {\bibinfo {author} {\bibfnamefont {G.}~\bibnamefont
  {Colangelo}}, \bibinfo {author} {\bibfnamefont {M.}~\bibnamefont
  {Hoferichter}},\ and\ \bibinfo {author} {\bibfnamefont {P.}~\bibnamefont
  {Stoffer}},\ }\href {https://doi.org/10.1007/JHEP02(2019)006} {\bibfield
  {journal} {\bibinfo  {journal} {JHEP}\ }\textbf {\bibinfo {volume} {02}},\
  \bibinfo {pages} {006}},\ \Eprint {https://arxiv.org/abs/1810.00007}
  {arXiv:1810.00007 [hep-ph]} \BibitemShut {NoStop}%
\bibitem [{\citenamefont {Hoferichter}\ \emph {et~al.}(2019)\citenamefont
  {Hoferichter}, \citenamefont {Hoid},\ and\ \citenamefont
  {Kubis}}]{hoferichter:2019mqg}%
  \BibitemOpen
  \bibfield  {author} {\bibinfo {author} {\bibfnamefont {M.}~\bibnamefont
  {Hoferichter}}, \bibinfo {author} {\bibfnamefont {B.-L.}\ \bibnamefont
  {Hoid}},\ and\ \bibinfo {author} {\bibfnamefont {B.}~\bibnamefont {Kubis}},\
  }\href {https://doi.org/10.1007/JHEP08(2019)137} {\bibfield  {journal}
  {\bibinfo  {journal} {JHEP}\ }\textbf {\bibinfo {volume} {08}},\ \bibinfo
  {pages} {137}},\ \Eprint {https://arxiv.org/abs/1907.01556} {arXiv:1907.01556
  [hep-ph]} \BibitemShut {NoStop}%
\bibitem [{\citenamefont {Davier}\ \emph {et~al.}(2020)\citenamefont {Davier},
  \citenamefont {Hoecker}, \citenamefont {Malaescu},\ and\ \citenamefont
  {Zhang}}]{davier:2019can}%
  \BibitemOpen
  \bibfield  {author} {\bibinfo {author} {\bibfnamefont {M.}~\bibnamefont
  {Davier}}, \bibinfo {author} {\bibfnamefont {A.}~\bibnamefont {Hoecker}},
  \bibinfo {author} {\bibfnamefont {B.}~\bibnamefont {Malaescu}},\ and\
  \bibinfo {author} {\bibfnamefont {Z.}~\bibnamefont {Zhang}},\ }\href
  {https://doi.org/10.1140/epjc/s10052-020-7792-2} {\bibfield  {journal}
  {\bibinfo  {journal} {Eur. Phys. J.}\ }\textbf {\bibinfo {volume} {C80}},\
  \bibinfo {pages} {241} (\bibinfo {year} {2020})},\ \bibinfo {note} {[Erratum:
  Eur. Phys. J. {\bf C80}, 410 (2020)]},\ \Eprint
  {https://arxiv.org/abs/1908.00921} {arXiv:1908.00921 [hep-ph]} \BibitemShut
  {NoStop}%
\bibitem [{\citenamefont {Keshavarzi}\ \emph
  {et~al.}(2020{\natexlab{a}})\citenamefont {Keshavarzi}, \citenamefont
  {Nomura},\ and\ \citenamefont {Teubner}}]{keshavarzi:2019abf}%
  \BibitemOpen
  \bibfield  {author} {\bibinfo {author} {\bibfnamefont {A.}~\bibnamefont
  {Keshavarzi}}, \bibinfo {author} {\bibfnamefont {D.}~\bibnamefont {Nomura}},\
  and\ \bibinfo {author} {\bibfnamefont {T.}~\bibnamefont {Teubner}},\ }\href
  {https://doi.org/10.1103/PhysRevD.101.014029} {\bibfield  {journal} {\bibinfo
   {journal} {Phys. Rev.}\ }\textbf {\bibinfo {volume} {D101}},\ \bibinfo
  {pages} {014029} (\bibinfo {year} {2020}{\natexlab{a}})},\ \Eprint
  {https://arxiv.org/abs/1911.00367} {arXiv:1911.00367 [hep-ph]} \BibitemShut
  {NoStop}%
\bibitem [{\citenamefont {Kurz}\ \emph {et~al.}(2014)\citenamefont {Kurz},
  \citenamefont {Liu}, \citenamefont {Marquard},\ and\ \citenamefont
  {Steinhauser}}]{kurz:2014wya}%
  \BibitemOpen
  \bibfield  {author} {\bibinfo {author} {\bibfnamefont {A.}~\bibnamefont
  {Kurz}}, \bibinfo {author} {\bibfnamefont {T.}~\bibnamefont {Liu}}, \bibinfo
  {author} {\bibfnamefont {P.}~\bibnamefont {Marquard}},\ and\ \bibinfo
  {author} {\bibfnamefont {M.}~\bibnamefont {Steinhauser}},\ }\href
  {https://doi.org/10.1016/j.physletb.2014.05.043} {\bibfield  {journal}
  {\bibinfo  {journal} {Phys. Lett.}\ }\textbf {\bibinfo {volume} {B734}},\
  \bibinfo {pages} {144} (\bibinfo {year} {2014})},\ \Eprint
  {https://arxiv.org/abs/1403.6400} {arXiv:1403.6400 [hep-ph]} \BibitemShut
  {NoStop}%
\bibitem [{\citenamefont {Melnikov}\ and\ \citenamefont
  {Vainshtein}(2004)}]{melnikov:2003xd}%
  \BibitemOpen
  \bibfield  {author} {\bibinfo {author} {\bibfnamefont {K.}~\bibnamefont
  {Melnikov}}\ and\ \bibinfo {author} {\bibfnamefont {A.}~\bibnamefont
  {Vainshtein}},\ }\href {https://doi.org/10.1103/PhysRevD.70.113006}
  {\bibfield  {journal} {\bibinfo  {journal} {Phys. Rev.}\ }\textbf {\bibinfo
  {volume} {D70}},\ \bibinfo {pages} {113006} (\bibinfo {year} {2004})},\
  \Eprint {https://arxiv.org/abs/hep-ph/0312226} {arXiv:hep-ph/0312226
  [hep-ph]} \BibitemShut {NoStop}%
\bibitem [{\citenamefont {Masjuan}\ and\ \citenamefont
  {S{\'a}nchez-Puertas}(2017)}]{masjuan:2017tvw}%
  \BibitemOpen
  \bibfield  {author} {\bibinfo {author} {\bibfnamefont {P.}~\bibnamefont
  {Masjuan}}\ and\ \bibinfo {author} {\bibfnamefont {P.}~\bibnamefont
  {S{\'a}nchez-Puertas}},\ }\href {https://doi.org/10.1103/PhysRevD.95.054026}
  {\bibfield  {journal} {\bibinfo  {journal} {Phys. Rev.}\ }\textbf {\bibinfo
  {volume} {D95}},\ \bibinfo {pages} {054026} (\bibinfo {year} {2017})},\
  \Eprint {https://arxiv.org/abs/1701.05829} {arXiv:1701.05829 [hep-ph]}
  \BibitemShut {NoStop}%
\bibitem [{\citenamefont {Colangelo}\ \emph {et~al.}(2017)\citenamefont
  {Colangelo}, \citenamefont {Hoferichter}, \citenamefont {Procura},\ and\
  \citenamefont {Stoffer}}]{Colangelo:2017fiz}%
  \BibitemOpen
  \bibfield  {author} {\bibinfo {author} {\bibfnamefont {G.}~\bibnamefont
  {Colangelo}}, \bibinfo {author} {\bibfnamefont {M.}~\bibnamefont
  {Hoferichter}}, \bibinfo {author} {\bibfnamefont {M.}~\bibnamefont
  {Procura}},\ and\ \bibinfo {author} {\bibfnamefont {P.}~\bibnamefont
  {Stoffer}},\ }\href {https://doi.org/10.1007/JHEP04(2017)161} {\bibfield
  {journal} {\bibinfo  {journal} {JHEP}\ }\textbf {\bibinfo {volume} {04}},\
  \bibinfo {pages} {161}},\ \Eprint {https://arxiv.org/abs/1702.07347}
  {arXiv:1702.07347 [hep-ph]} \BibitemShut {NoStop}%
\bibitem [{\citenamefont {Hoferichter}\ \emph {et~al.}(2018)\citenamefont
  {Hoferichter}, \citenamefont {Hoid}, \citenamefont {Kubis}, \citenamefont
  {Leupold},\ and\ \citenamefont {Schneider}}]{hoferichter:2018kwz}%
  \BibitemOpen
  \bibfield  {author} {\bibinfo {author} {\bibfnamefont {M.}~\bibnamefont
  {Hoferichter}}, \bibinfo {author} {\bibfnamefont {B.-L.}\ \bibnamefont
  {Hoid}}, \bibinfo {author} {\bibfnamefont {B.}~\bibnamefont {Kubis}},
  \bibinfo {author} {\bibfnamefont {S.}~\bibnamefont {Leupold}},\ and\ \bibinfo
  {author} {\bibfnamefont {S.~P.}\ \bibnamefont {Schneider}},\ }\href
  {https://doi.org/10.1007/JHEP10(2018)141} {\bibfield  {journal} {\bibinfo
  {journal} {JHEP}\ }\textbf {\bibinfo {volume} {10}},\ \bibinfo {pages}
  {141}},\ \Eprint {https://arxiv.org/abs/1808.04823} {arXiv:1808.04823
  [hep-ph]} \BibitemShut {NoStop}%
\bibitem [{\citenamefont {G{\'e}rardin}\ \emph {et~al.}(2019)\citenamefont
  {G{\'e}rardin}, \citenamefont {Meyer},\ and\ \citenamefont
  {Nyffeler}}]{gerardin:2019vio}%
  \BibitemOpen
  \bibfield  {author} {\bibinfo {author} {\bibfnamefont {A.}~\bibnamefont
  {G{\'e}rardin}}, \bibinfo {author} {\bibfnamefont {H.~B.}\ \bibnamefont
  {Meyer}},\ and\ \bibinfo {author} {\bibfnamefont {A.}~\bibnamefont
  {Nyffeler}},\ }\href {https://doi.org/10.1103/PhysRevD.100.034520} {\bibfield
   {journal} {\bibinfo  {journal} {Phys. Rev.}\ }\textbf {\bibinfo {volume}
  {D100}},\ \bibinfo {pages} {034520} (\bibinfo {year} {2019})},\ \Eprint
  {https://arxiv.org/abs/1903.09471} {arXiv:1903.09471 [hep-lat]} \BibitemShut
  {NoStop}%
\bibitem [{\citenamefont {Bijnens}\ \emph {et~al.}(2019)\citenamefont
  {Bijnens}, \citenamefont {Hermansson-Truedsson},\ and\ \citenamefont
  {Rodr{\'i}guez-S{\'a}nchez}}]{bijnens:2019ghy}%
  \BibitemOpen
  \bibfield  {author} {\bibinfo {author} {\bibfnamefont {J.}~\bibnamefont
  {Bijnens}}, \bibinfo {author} {\bibfnamefont {N.}~\bibnamefont
  {Hermansson-Truedsson}},\ and\ \bibinfo {author} {\bibfnamefont
  {A.}~\bibnamefont {Rodr{\'i}guez-S{\'a}nchez}},\ }\href
  {https://doi.org/10.1016/j.physletb.2019.134994} {\bibfield  {journal}
  {\bibinfo  {journal} {Phys. Lett.}\ }\textbf {\bibinfo {volume} {B798}},\
  \bibinfo {pages} {134994} (\bibinfo {year} {2019})},\ \Eprint
  {https://arxiv.org/abs/1908.03331} {arXiv:1908.03331 [hep-ph]} \BibitemShut
  {NoStop}%
\bibitem [{\citenamefont {Colangelo}\ \emph {et~al.}(2020)\citenamefont
  {Colangelo}, \citenamefont {Hagelstein}, \citenamefont {Hoferichter},
  \citenamefont {Laub},\ and\ \citenamefont {Stoffer}}]{colangelo:2019uex}%
  \BibitemOpen
  \bibfield  {author} {\bibinfo {author} {\bibfnamefont {G.}~\bibnamefont
  {Colangelo}}, \bibinfo {author} {\bibfnamefont {F.}~\bibnamefont
  {Hagelstein}}, \bibinfo {author} {\bibfnamefont {M.}~\bibnamefont
  {Hoferichter}}, \bibinfo {author} {\bibfnamefont {L.}~\bibnamefont {Laub}},\
  and\ \bibinfo {author} {\bibfnamefont {P.}~\bibnamefont {Stoffer}},\ }\href
  {https://doi.org/10.1007/JHEP03(2020)101} {\bibfield  {journal} {\bibinfo
  {journal} {JHEP}\ }\textbf {\bibinfo {volume} {03}},\ \bibinfo {pages}
  {101}},\ \Eprint {https://arxiv.org/abs/1910.13432} {arXiv:1910.13432
  [hep-ph]} \BibitemShut {NoStop}%
\bibitem [{\citenamefont {Blum}\ \emph {et~al.}(2020)\citenamefont {Blum},
  \citenamefont {Christ}, \citenamefont {Hayakawa}, \citenamefont {Izubuchi},
  \citenamefont {Jin}, \citenamefont {Jung},\ and\ \citenamefont
  {Lehner}}]{Blum:2019ugy}%
  \BibitemOpen
  \bibfield  {author} {\bibinfo {author} {\bibfnamefont {T.}~\bibnamefont
  {Blum}}, \bibinfo {author} {\bibfnamefont {N.}~\bibnamefont {Christ}},
  \bibinfo {author} {\bibfnamefont {M.}~\bibnamefont {Hayakawa}}, \bibinfo
  {author} {\bibfnamefont {T.}~\bibnamefont {Izubuchi}}, \bibinfo {author}
  {\bibfnamefont {L.}~\bibnamefont {Jin}}, \bibinfo {author} {\bibfnamefont
  {C.}~\bibnamefont {Jung}},\ and\ \bibinfo {author} {\bibfnamefont
  {C.}~\bibnamefont {Lehner}},\ }\href
  {https://doi.org/10.1103/PhysRevLett.124.132002} {\bibfield  {journal}
  {\bibinfo  {journal} {Phys. Rev. Lett.}\ }\textbf {\bibinfo {volume} {124}},\
  \bibinfo {pages} {132002} (\bibinfo {year} {2020})},\ \Eprint
  {https://arxiv.org/abs/1911.08123} {arXiv:1911.08123 [hep-lat]} \BibitemShut
  {NoStop}%
\bibitem [{\citenamefont {Colangelo}\ \emph {et~al.}(2014)\citenamefont
  {Colangelo}, \citenamefont {Hoferichter}, \citenamefont {Nyffeler},
  \citenamefont {Passera},\ and\ \citenamefont {Stoffer}}]{colangelo:2014qya}%
  \BibitemOpen
  \bibfield  {author} {\bibinfo {author} {\bibfnamefont {G.}~\bibnamefont
  {Colangelo}}, \bibinfo {author} {\bibfnamefont {M.}~\bibnamefont
  {Hoferichter}}, \bibinfo {author} {\bibfnamefont {A.}~\bibnamefont
  {Nyffeler}}, \bibinfo {author} {\bibfnamefont {M.}~\bibnamefont {Passera}},\
  and\ \bibinfo {author} {\bibfnamefont {P.}~\bibnamefont {Stoffer}},\ }\href
  {https://doi.org/10.1016/j.physletb.2014.06.012} {\bibfield  {journal}
  {\bibinfo  {journal} {Phys. Lett.}\ }\textbf {\bibinfo {volume} {B735}},\
  \bibinfo {pages} {90} (\bibinfo {year} {2014})},\ \Eprint
  {https://arxiv.org/abs/1403.7512} {arXiv:1403.7512 [hep-ph]} \BibitemShut
  {NoStop}%
\bibitem [{\citenamefont {Brodsky}\ and\ \citenamefont
  {De~Rafael}(1968)}]{Brodsky:1967sr}%
  \BibitemOpen
  \bibfield  {author} {\bibinfo {author} {\bibfnamefont {S.~J.}\ \bibnamefont
  {Brodsky}}\ and\ \bibinfo {author} {\bibfnamefont {E.}~\bibnamefont
  {De~Rafael}},\ }\href {https://doi.org/10.1103/PhysRev.168.1620} {\bibfield
  {journal} {\bibinfo  {journal} {Phys. Rev.}\ }\textbf {\bibinfo {volume}
  {168}},\ \bibinfo {pages} {1620} (\bibinfo {year} {1968})}\BibitemShut
  {NoStop}%
\bibitem [{\citenamefont {Lautrup}\ and\ \citenamefont
  {De~Rafael}(1968)}]{Lautrup:1968tdb}%
  \BibitemOpen
  \bibfield  {author} {\bibinfo {author} {\bibfnamefont {B.~E.}\ \bibnamefont
  {Lautrup}}\ and\ \bibinfo {author} {\bibfnamefont {E.}~\bibnamefont
  {De~Rafael}},\ }\href {https://doi.org/10.1103/PhysRev.174.1835} {\bibfield
  {journal} {\bibinfo  {journal} {Phys. Rev.}\ }\textbf {\bibinfo {volume}
  {174}},\ \bibinfo {pages} {1835} (\bibinfo {year} {1968})}\BibitemShut
  {NoStop}%
\bibitem [{\citenamefont {Krause}(1997)}]{Krause:1996rf}%
  \BibitemOpen
  \bibfield  {author} {\bibinfo {author} {\bibfnamefont {B.}~\bibnamefont
  {Krause}},\ }\href {https://doi.org/10.1016/S0370-2693(96)01346-9} {\bibfield
   {journal} {\bibinfo  {journal} {Phys. Lett. B}\ }\textbf {\bibinfo {volume}
  {390}},\ \bibinfo {pages} {392} (\bibinfo {year} {1997})},\ \Eprint
  {https://arxiv.org/abs/hep-ph/9607259} {arXiv:hep-ph/9607259} \BibitemShut
  {NoStop}%
\bibitem [{\citenamefont {Jegerlehner}(2017)}]{Jegerlehner:2017gek}%
  \BibitemOpen
  \bibfield  {author} {\bibinfo {author} {\bibfnamefont {F.}~\bibnamefont
  {Jegerlehner}},\ }\href {https://doi.org/10.1007/978-3-319-63577-4} {}Vol.\
  \bibinfo {volume} {274}\ (\bibinfo  {publisher} {Springer},\ \bibinfo
  {address} {Cham},\ \bibinfo {year} {2017})\BibitemShut {NoStop}%
\bibitem [{\citenamefont {Jegerlehner}(2016)}]{Jegerlehner:2015stw}%
  \BibitemOpen
  \bibfield  {author} {\bibinfo {author} {\bibfnamefont {F.}~\bibnamefont
  {Jegerlehner}},\ }\href {https://doi.org/10.1051/epjconf/201611801016}
  {\bibfield  {journal} {\bibinfo  {journal} {EPJ Web Conf.}\ }\textbf
  {\bibinfo {volume} {118}},\ \bibinfo {pages} {01016} (\bibinfo {year}
  {2016})},\ \Eprint {https://arxiv.org/abs/1511.04473} {arXiv:1511.04473
  [hep-ph]} \BibitemShut {NoStop}%
\bibitem [{\citenamefont {Jegerlehner}(2018)}]{Jegerlehner:2017lbd}%
  \BibitemOpen
  \bibfield  {author} {\bibinfo {author} {\bibfnamefont {F.}~\bibnamefont
  {Jegerlehner}},\ }\href {https://doi.org/10.1051/epjconf/201816600022}
  {\bibfield  {journal} {\bibinfo  {journal} {EPJ Web Conf.}\ }\textbf
  {\bibinfo {volume} {166}},\ \bibinfo {pages} {00022} (\bibinfo {year}
  {2018})},\ \Eprint {https://arxiv.org/abs/1705.00263} {arXiv:1705.00263
  [hep-ph]} \BibitemShut {NoStop}%
\bibitem [{\citenamefont
  {Jegerlehner}(2019{\natexlab{a}})}]{Jegerlehner:2017zsb}%
  \BibitemOpen
  \bibfield  {author} {\bibinfo {author} {\bibfnamefont {F.}~\bibnamefont
  {Jegerlehner}},\ }\href {https://doi.org/10.1051/epjconf/201921801003}
  {\bibfield  {journal} {\bibinfo  {journal} {EPJ Web Conf.}\ }\textbf
  {\bibinfo {volume} {218}},\ \bibinfo {pages} {01003} (\bibinfo {year}
  {2019}{\natexlab{a}})},\ \Eprint {https://arxiv.org/abs/1711.06089}
  {arXiv:1711.06089 [hep-ph]} \BibitemShut {NoStop}%
\bibitem [{\citenamefont
  {Jegerlehner}(2019{\natexlab{b}})}]{Jegerlehner:2018gjd}%
  \BibitemOpen
  \bibfield  {author} {\bibinfo {author} {\bibfnamefont {F.}~\bibnamefont
  {Jegerlehner}},\ }\href {https://doi.org/10.1051/epjconf/201919901010}
  {\bibfield  {journal} {\bibinfo  {journal} {EPJ Web Conf.}\ }\textbf
  {\bibinfo {volume} {199}},\ \bibinfo {pages} {01010} (\bibinfo {year}
  {2019}{\natexlab{b}})},\ \Eprint {https://arxiv.org/abs/1809.07413}
  {arXiv:1809.07413 [hep-ph]} \BibitemShut {NoStop}%
\bibitem [{\citenamefont {Eidelman}\ and\ \citenamefont
  {Jegerlehner}(1995)}]{Eidelman:1995ny}%
  \BibitemOpen
  \bibfield  {author} {\bibinfo {author} {\bibfnamefont {S.}~\bibnamefont
  {Eidelman}}\ and\ \bibinfo {author} {\bibfnamefont {F.}~\bibnamefont
  {Jegerlehner}},\ }\href {https://doi.org/10.1007/BF01553984} {\bibfield
  {journal} {\bibinfo  {journal} {Z. Phys. C}\ }\textbf {\bibinfo {volume}
  {67}},\ \bibinfo {pages} {585} (\bibinfo {year} {1995})},\ \Eprint
  {https://arxiv.org/abs/hep-ph/9502298} {arXiv:hep-ph/9502298} \BibitemShut
  {NoStop}%
\bibitem [{\citenamefont {Benayoun}\ \emph {et~al.}(2008)\citenamefont
  {Benayoun}, \citenamefont {David}, \citenamefont {DelBuono}, \citenamefont
  {Leitner},\ and\ \citenamefont {O'Connell}}]{Benayoun:2007cu}%
  \BibitemOpen
  \bibfield  {author} {\bibinfo {author} {\bibfnamefont {M.}~\bibnamefont
  {Benayoun}}, \bibinfo {author} {\bibfnamefont {P.}~\bibnamefont {David}},
  \bibinfo {author} {\bibfnamefont {L.}~\bibnamefont {DelBuono}}, \bibinfo
  {author} {\bibfnamefont {O.}~\bibnamefont {Leitner}},\ and\ \bibinfo {author}
  {\bibfnamefont {H.~B.}\ \bibnamefont {O'Connell}},\ }\href
  {https://doi.org/10.1140/epjc/s10052-008-0586-6} {\bibfield  {journal}
  {\bibinfo  {journal} {Eur. Phys. J. C}\ }\textbf {\bibinfo {volume} {55}},\
  \bibinfo {pages} {199} (\bibinfo {year} {2008})},\ \Eprint
  {https://arxiv.org/abs/0711.4482} {arXiv:0711.4482 [hep-ph]} \BibitemShut
  {NoStop}%
\bibitem [{\citenamefont {Benayoun}\ \emph {et~al.}(2012)\citenamefont
  {Benayoun}, \citenamefont {David}, \citenamefont {DelBuono},\ and\
  \citenamefont {Jegerlehner}}]{Benayoun:2012etq}%
  \BibitemOpen
  \bibfield  {author} {\bibinfo {author} {\bibfnamefont {M.}~\bibnamefont
  {Benayoun}}, \bibinfo {author} {\bibfnamefont {P.}~\bibnamefont {David}},
  \bibinfo {author} {\bibfnamefont {L.}~\bibnamefont {DelBuono}},\ and\
  \bibinfo {author} {\bibfnamefont {F.}~\bibnamefont {Jegerlehner}},\ }\href
  {https://doi.org/10.1140/epjc/s10052-011-1848-2} {\bibfield  {journal}
  {\bibinfo  {journal} {Eur. Phys. J. C}\ }\textbf {\bibinfo {volume} {72}},\
  \bibinfo {pages} {1848} (\bibinfo {year} {2012})},\ \Eprint
  {https://arxiv.org/abs/1106.1315} {arXiv:1106.1315 [hep-ph]} \BibitemShut
  {NoStop}%
\bibitem [{\citenamefont {Benayoun}\ \emph {et~al.}(2013)\citenamefont
  {Benayoun}, \citenamefont {David}, \citenamefont {DelBuono},\ and\
  \citenamefont {Jegerlehner}}]{Benayoun:2012wc}%
  \BibitemOpen
  \bibfield  {author} {\bibinfo {author} {\bibfnamefont {M.}~\bibnamefont
  {Benayoun}}, \bibinfo {author} {\bibfnamefont {P.}~\bibnamefont {David}},
  \bibinfo {author} {\bibfnamefont {L.}~\bibnamefont {DelBuono}},\ and\
  \bibinfo {author} {\bibfnamefont {F.}~\bibnamefont {Jegerlehner}},\ }\href
  {https://doi.org/10.1140/epjc/s10052-013-2453-3} {\bibfield  {journal}
  {\bibinfo  {journal} {Eur. Phys. J. C}\ }\textbf {\bibinfo {volume} {73}},\
  \bibinfo {pages} {2453} (\bibinfo {year} {2013})},\ \Eprint
  {https://arxiv.org/abs/1210.7184} {arXiv:1210.7184 [hep-ph]} \BibitemShut
  {NoStop}%
\bibitem [{\citenamefont {Benayoun}\ \emph {et~al.}(2015)\citenamefont
  {Benayoun}, \citenamefont {David}, \citenamefont {DelBuono},\ and\
  \citenamefont {Jegerlehner}}]{Benayoun:2015gxa}%
  \BibitemOpen
  \bibfield  {author} {\bibinfo {author} {\bibfnamefont {M.}~\bibnamefont
  {Benayoun}}, \bibinfo {author} {\bibfnamefont {P.}~\bibnamefont {David}},
  \bibinfo {author} {\bibfnamefont {L.}~\bibnamefont {DelBuono}},\ and\
  \bibinfo {author} {\bibfnamefont {F.}~\bibnamefont {Jegerlehner}},\ }\href
  {https://doi.org/10.1140/epjc/s10052-015-3830-x} {\bibfield  {journal}
  {\bibinfo  {journal} {Eur. Phys. J. C}\ }\textbf {\bibinfo {volume} {75}},\
  \bibinfo {pages} {613} (\bibinfo {year} {2015})},\ \Eprint
  {https://arxiv.org/abs/1507.02943} {arXiv:1507.02943 [hep-ph]} \BibitemShut
  {NoStop}%
\bibitem [{\citenamefont {Benayoun}\ \emph {et~al.}(2020)\citenamefont
  {Benayoun}, \citenamefont {Delbuono},\ and\ \citenamefont
  {Jegerlehner}}]{Benayoun:2019zwh}%
  \BibitemOpen
  \bibfield  {author} {\bibinfo {author} {\bibfnamefont {M.}~\bibnamefont
  {Benayoun}}, \bibinfo {author} {\bibfnamefont {L.}~\bibnamefont {Delbuono}},\
  and\ \bibinfo {author} {\bibfnamefont {F.}~\bibnamefont {Jegerlehner}},\
  }\href {https://doi.org/10.1140/epjc/s10052-020-7611-9} {\bibfield  {journal}
  {\bibinfo  {journal} {Eur. Phys. J. C}\ }\textbf {\bibinfo {volume} {80}},\
  \bibinfo {pages} {81} (\bibinfo {year} {2020})},\ \bibinfo {note} {[Erratum:
  Eur.Phys.J.C 80, 244 (2020)]},\ \Eprint {https://arxiv.org/abs/1903.11034}
  {arXiv:1903.11034 [hep-ph]} \BibitemShut {NoStop}%
\bibitem [{\citenamefont {Davier}\ \emph {et~al.}(2011)\citenamefont {Davier},
  \citenamefont {Hoecker}, \citenamefont {Malaescu},\ and\ \citenamefont
  {Zhang}}]{Davier:2010nc}%
  \BibitemOpen
  \bibfield  {author} {\bibinfo {author} {\bibfnamefont {M.}~\bibnamefont
  {Davier}}, \bibinfo {author} {\bibfnamefont {A.}~\bibnamefont {Hoecker}},
  \bibinfo {author} {\bibfnamefont {B.}~\bibnamefont {Malaescu}},\ and\
  \bibinfo {author} {\bibfnamefont {Z.}~\bibnamefont {Zhang}},\ }\href
  {https://doi.org/10.1140/epjc/s10052-012-1874-8} {\bibfield  {journal}
  {\bibinfo  {journal} {Eur. Phys. J. C}\ }\textbf {\bibinfo {volume} {71}},\
  \bibinfo {pages} {1515} (\bibinfo {year} {2011})},\ \bibinfo {note}
  {[Erratum: Eur.Phys.J.C 72, 1874 (2012)]},\ \Eprint
  {https://arxiv.org/abs/1010.4180} {arXiv:1010.4180 [hep-ph]} \BibitemShut
  {NoStop}%
\bibitem [{\citenamefont {Borsanyi}\ \emph {et~al.}(2018)\citenamefont
  {Borsanyi} \emph {et~al.}}]{Budapest-Marseille-Wuppertal:2017okr}%
  \BibitemOpen
  \bibfield  {author} {\bibinfo {author} {\bibfnamefont {S.}~\bibnamefont
  {Borsanyi}} \emph {et~al.} (\bibinfo {collaboration}
  {Budapest-Marseille-Wuppertal}),\ }\href
  {https://doi.org/10.1103/PhysRevLett.121.022002} {\bibfield  {journal}
  {\bibinfo  {journal} {Phys. Rev. Lett.}\ }\textbf {\bibinfo {volume} {121}},\
  \bibinfo {pages} {022002} (\bibinfo {year} {2018})},\ \Eprint
  {https://arxiv.org/abs/1711.04980} {arXiv:1711.04980 [hep-lat]} \BibitemShut
  {NoStop}%
\bibitem [{\citenamefont {Blum}\ \emph
  {et~al.}(2018{\natexlab{a}})\citenamefont {Blum}, \citenamefont {Boyle},
  \citenamefont {G\"ulpers}, \citenamefont {Izubuchi}, \citenamefont {Jin},
  \citenamefont {Jung}, \citenamefont {J\"uttner}, \citenamefont {Lehner},
  \citenamefont {Portelli},\ and\ \citenamefont {Tsang}}]{RBC:2018dos}%
  \BibitemOpen
  \bibfield  {author} {\bibinfo {author} {\bibfnamefont {T.}~\bibnamefont
  {Blum}}, \bibinfo {author} {\bibfnamefont {P.~A.}\ \bibnamefont {Boyle}},
  \bibinfo {author} {\bibfnamefont {V.}~\bibnamefont {G\"ulpers}}, \bibinfo
  {author} {\bibfnamefont {T.}~\bibnamefont {Izubuchi}}, \bibinfo {author}
  {\bibfnamefont {L.}~\bibnamefont {Jin}}, \bibinfo {author} {\bibfnamefont
  {C.}~\bibnamefont {Jung}}, \bibinfo {author} {\bibfnamefont {A.}~\bibnamefont
  {J\"uttner}}, \bibinfo {author} {\bibfnamefont {C.}~\bibnamefont {Lehner}},
  \bibinfo {author} {\bibfnamefont {A.}~\bibnamefont {Portelli}},\ and\
  \bibinfo {author} {\bibfnamefont {J.~T.}\ \bibnamefont {Tsang}} (\bibinfo
  {collaboration} {RBC, UKQCD}),\ }\href
  {https://doi.org/10.1103/PhysRevLett.121.022003} {\bibfield  {journal}
  {\bibinfo  {journal} {Phys. Rev. Lett.}\ }\textbf {\bibinfo {volume} {121}},\
  \bibinfo {pages} {022003} (\bibinfo {year} {2018}{\natexlab{a}})},\ \Eprint
  {https://arxiv.org/abs/1801.07224} {arXiv:1801.07224 [hep-lat]} \BibitemShut
  {NoStop}%
\bibitem [{\citenamefont {Giusti}\ \emph {et~al.}(2019)\citenamefont {Giusti},
  \citenamefont {Lubicz}, \citenamefont {Martinelli}, \citenamefont
  {Sanfilippo},\ and\ \citenamefont {Simula}}]{Giusti:2019xct}%
  \BibitemOpen
  \bibfield  {author} {\bibinfo {author} {\bibfnamefont {D.}~\bibnamefont
  {Giusti}}, \bibinfo {author} {\bibfnamefont {V.}~\bibnamefont {Lubicz}},
  \bibinfo {author} {\bibfnamefont {G.}~\bibnamefont {Martinelli}}, \bibinfo
  {author} {\bibfnamefont {F.}~\bibnamefont {Sanfilippo}},\ and\ \bibinfo
  {author} {\bibfnamefont {S.}~\bibnamefont {Simula}},\ }\href
  {https://doi.org/10.1103/PhysRevD.99.114502} {\bibfield  {journal} {\bibinfo
  {journal} {Phys. Rev. D}\ }\textbf {\bibinfo {volume} {99}},\ \bibinfo
  {pages} {114502} (\bibinfo {year} {2019})},\ \Eprint
  {https://arxiv.org/abs/1901.10462} {arXiv:1901.10462 [hep-lat]} \BibitemShut
  {NoStop}%
\bibitem [{\citenamefont {Davies}\ \emph {et~al.}(2020)\citenamefont {Davies}
  \emph {et~al.}}]{FermilabLattice:2019ugu}%
  \BibitemOpen
  \bibfield  {author} {\bibinfo {author} {\bibfnamefont {C.~T.~H.}\
  \bibnamefont {Davies}} \emph {et~al.} (\bibinfo {collaboration} {Fermilab
  Lattice, LATTICE-HPQCD, MILC}),\ }\href
  {https://doi.org/10.1103/PhysRevD.101.034512} {\bibfield  {journal} {\bibinfo
   {journal} {Phys. Rev. D}\ }\textbf {\bibinfo {volume} {101}},\ \bibinfo
  {pages} {034512} (\bibinfo {year} {2020})},\ \Eprint
  {https://arxiv.org/abs/1902.04223} {arXiv:1902.04223 [hep-lat]} \BibitemShut
  {NoStop}%
\bibitem [{\citenamefont {G\'erardin}\ \emph {et~al.}(2019)\citenamefont
  {G\'erardin}, \citenamefont {C\`e}, \citenamefont {von Hippel}, \citenamefont
  {H\"orz}, \citenamefont {Meyer}, \citenamefont {Mohler}, \citenamefont
  {Ottnad}, \citenamefont {Wilhelm},\ and\ \citenamefont
  {Wittig}}]{Gerardin:2019rua}%
  \BibitemOpen
  \bibfield  {author} {\bibinfo {author} {\bibfnamefont {A.}~\bibnamefont
  {G\'erardin}}, \bibinfo {author} {\bibfnamefont {M.}~\bibnamefont {C\`e}},
  \bibinfo {author} {\bibfnamefont {G.}~\bibnamefont {von Hippel}}, \bibinfo
  {author} {\bibfnamefont {B.}~\bibnamefont {H\"orz}}, \bibinfo {author}
  {\bibfnamefont {H.~B.}\ \bibnamefont {Meyer}}, \bibinfo {author}
  {\bibfnamefont {D.}~\bibnamefont {Mohler}}, \bibinfo {author} {\bibfnamefont
  {K.}~\bibnamefont {Ottnad}}, \bibinfo {author} {\bibfnamefont
  {J.}~\bibnamefont {Wilhelm}},\ and\ \bibinfo {author} {\bibfnamefont
  {H.}~\bibnamefont {Wittig}},\ }\href
  {https://doi.org/10.1103/PhysRevD.100.014510} {\bibfield  {journal} {\bibinfo
   {journal} {Phys. Rev. D}\ }\textbf {\bibinfo {volume} {100}},\ \bibinfo
  {pages} {014510} (\bibinfo {year} {2019})},\ \Eprint
  {https://arxiv.org/abs/1904.03120} {arXiv:1904.03120 [hep-lat]} \BibitemShut
  {NoStop}%
\bibitem [{\citenamefont {Chakraborty}\ \emph {et~al.}(2018)\citenamefont
  {Chakraborty} \emph {et~al.}}]{chakraborty:2017tqp}%
  \BibitemOpen
  \bibfield  {author} {\bibinfo {author} {\bibfnamefont {B.}~\bibnamefont
  {Chakraborty}} \emph {et~al.} (\bibinfo {collaboration} {Fermilab Lattice,
  LATTICE-HPQCD, MILC}),\ }\href
  {https://doi.org/10.1103/PhysRevLett.120.152001} {\bibfield  {journal}
  {\bibinfo  {journal} {Phys. Rev. Lett.}\ }\textbf {\bibinfo {volume} {120}},\
  \bibinfo {pages} {152001} (\bibinfo {year} {2018})},\ \Eprint
  {https://arxiv.org/abs/1710.11212} {arXiv:1710.11212 [hep-lat]} \BibitemShut
  {NoStop}%
\bibitem [{\citenamefont {Blum}\ \emph
  {et~al.}(2018{\natexlab{b}})\citenamefont {Blum}, \citenamefont {Boyle},
  \citenamefont {G{\"u}lpers}, \citenamefont {Izubuchi}, \citenamefont {Jin},
  \citenamefont {Jung}, \citenamefont {J{\"u}ttner}, \citenamefont {Lehner},
  \citenamefont {Portelli},\ and\ \citenamefont {Tsang}}]{blum:2018mom}%
  \BibitemOpen
  \bibfield  {author} {\bibinfo {author} {\bibfnamefont {T.}~\bibnamefont
  {Blum}}, \bibinfo {author} {\bibfnamefont {P.~A.}\ \bibnamefont {Boyle}},
  \bibinfo {author} {\bibfnamefont {V.}~\bibnamefont {G{\"u}lpers}}, \bibinfo
  {author} {\bibfnamefont {T.}~\bibnamefont {Izubuchi}}, \bibinfo {author}
  {\bibfnamefont {L.}~\bibnamefont {Jin}}, \bibinfo {author} {\bibfnamefont
  {C.}~\bibnamefont {Jung}}, \bibinfo {author} {\bibfnamefont {A.}~\bibnamefont
  {J{\"u}ttner}}, \bibinfo {author} {\bibfnamefont {C.}~\bibnamefont {Lehner}},
  \bibinfo {author} {\bibfnamefont {A.}~\bibnamefont {Portelli}},\ and\
  \bibinfo {author} {\bibfnamefont {J.~T.}\ \bibnamefont {Tsang}} (\bibinfo
  {collaboration} {RBC, UKQCD}),\ }\href
  {https://doi.org/10.1103/PhysRevLett.121.022003} {\bibfield  {journal}
  {\bibinfo  {journal} {Phys. Rev. Lett.}\ }\textbf {\bibinfo {volume} {121}},\
  \bibinfo {pages} {022003} (\bibinfo {year} {2018}{\natexlab{b}})},\ \Eprint
  {https://arxiv.org/abs/1801.07224} {arXiv:1801.07224 [hep-lat]} \BibitemShut
  {NoStop}%
\bibitem [{\citenamefont {Shintani}\ and\ \citenamefont
  {Kuramashi}(2019)}]{shintani:2019wai}%
  \BibitemOpen
  \bibfield  {author} {\bibinfo {author} {\bibfnamefont {E.}~\bibnamefont
  {Shintani}}\ and\ \bibinfo {author} {\bibfnamefont {Y.}~\bibnamefont
  {Kuramashi}},\ }\href {https://doi.org/10.1103/PhysRevD.100.034517}
  {\bibfield  {journal} {\bibinfo  {journal} {Phys. Rev.}\ }\textbf {\bibinfo
  {volume} {D100}},\ \bibinfo {pages} {034517} (\bibinfo {year} {2019})},\
  \Eprint {https://arxiv.org/abs/1902.00885} {arXiv:1902.00885 [hep-lat]}
  \BibitemShut {NoStop}%
\bibitem [{\citenamefont {Aubin}\ \emph {et~al.}(2020)\citenamefont {Aubin},
  \citenamefont {Blum}, \citenamefont {Tu}, \citenamefont {Golterman},
  \citenamefont {Jung},\ and\ \citenamefont {Peris}}]{Aubin:2019usy}%
  \BibitemOpen
  \bibfield  {author} {\bibinfo {author} {\bibfnamefont {C.}~\bibnamefont
  {Aubin}}, \bibinfo {author} {\bibfnamefont {T.}~\bibnamefont {Blum}},
  \bibinfo {author} {\bibfnamefont {C.}~\bibnamefont {Tu}}, \bibinfo {author}
  {\bibfnamefont {M.}~\bibnamefont {Golterman}}, \bibinfo {author}
  {\bibfnamefont {C.}~\bibnamefont {Jung}},\ and\ \bibinfo {author}
  {\bibfnamefont {S.}~\bibnamefont {Peris}},\ }\href
  {https://doi.org/10.1103/PhysRevD.101.014503} {\bibfield  {journal} {\bibinfo
   {journal} {Phys. Rev.}\ }\textbf {\bibinfo {volume} {D101}},\ \bibinfo
  {pages} {014503} (\bibinfo {year} {2020})},\ \Eprint
  {https://arxiv.org/abs/1905.09307} {arXiv:1905.09307 [hep-lat]} \BibitemShut
  {NoStop}%
\bibitem [{\citenamefont {Giusti}\ and\ \citenamefont
  {Simula}(2019)}]{giusti:2019hkz}%
  \BibitemOpen
  \bibfield  {author} {\bibinfo {author} {\bibfnamefont {D.}~\bibnamefont
  {Giusti}}\ and\ \bibinfo {author} {\bibfnamefont {S.}~\bibnamefont
  {Simula}},\ }\href {https://doi.org/10.22323/1.363.0104} {\bibfield
  {journal} {\bibinfo  {journal} {PoS}\ }\textbf {\bibinfo {volume}
  {LATTICE2019}},\ \bibinfo {pages} {104} (\bibinfo {year} {2019})},\ \Eprint
  {https://arxiv.org/abs/1910.03874} {arXiv:1910.03874 [hep-lat]} \BibitemShut
  {NoStop}%
\bibitem [{\citenamefont {Borsanyi}\ \emph {et~al.}(2021)\citenamefont
  {Borsanyi} \emph {et~al.}}]{Borsanyi:2020mff}%
  \BibitemOpen
  \bibfield  {author} {\bibinfo {author} {\bibfnamefont {S.}~\bibnamefont
  {Borsanyi}} \emph {et~al.},\ }\href
  {https://doi.org/10.1038/s41586-021-03418-1} {\bibfield  {journal} {\bibinfo
  {journal} {Nature}\ }\textbf {\bibinfo {volume} {593}},\ \bibinfo {pages}
  {51} (\bibinfo {year} {2021})},\ \Eprint {https://arxiv.org/abs/2002.12347}
  {arXiv:2002.12347 [hep-lat]} \BibitemShut {NoStop}%
\bibitem [{\citenamefont {Boccaletti}\ \emph {et~al.}(2024)\citenamefont
  {Boccaletti} \emph {et~al.}}]{Boccaletti:2024guq}%
  \BibitemOpen
  \bibfield  {author} {\bibinfo {author} {\bibfnamefont {A.}~\bibnamefont
  {Boccaletti}} \emph {et~al.},\ }\href@noop {} {\  (\bibinfo {year} {2024})},\
  \Eprint {https://arxiv.org/abs/2407.10913} {arXiv:2407.10913 [hep-lat]}
  \BibitemShut {NoStop}%
\bibitem [{\citenamefont {Lehner}\ and\ \citenamefont
  {Meyer}(2020)}]{Lehner:2020crt}%
  \BibitemOpen
  \bibfield  {author} {\bibinfo {author} {\bibfnamefont {C.}~\bibnamefont
  {Lehner}}\ and\ \bibinfo {author} {\bibfnamefont {A.~S.}\ \bibnamefont
  {Meyer}},\ }\href {https://doi.org/10.1103/PhysRevD.101.074515} {\bibfield
  {journal} {\bibinfo  {journal} {Phys. Rev. D}\ }\textbf {\bibinfo {volume}
  {101}},\ \bibinfo {pages} {074515} (\bibinfo {year} {2020})},\ \Eprint
  {https://arxiv.org/abs/2003.04177} {arXiv:2003.04177 [hep-lat]} \BibitemShut
  {NoStop}%
\bibitem [{\citenamefont {Alexandrou}\ \emph {et~al.}(2023)\citenamefont
  {Alexandrou} \emph {et~al.}}]{ExtendedTwistedMass:2022jpw}%
  \BibitemOpen
  \bibfield  {author} {\bibinfo {author} {\bibfnamefont {C.}~\bibnamefont
  {Alexandrou}} \emph {et~al.} (\bibinfo {collaboration} {Extended Twisted
  Mass}),\ }\href {https://doi.org/10.1103/PhysRevD.107.074506} {\bibfield
  {journal} {\bibinfo  {journal} {Phys. Rev. D}\ }\textbf {\bibinfo {volume}
  {107}},\ \bibinfo {pages} {074506} (\bibinfo {year} {2023})},\ \Eprint
  {https://arxiv.org/abs/2206.15084} {arXiv:2206.15084 [hep-lat]} \BibitemShut
  {NoStop}%
\bibitem [{\citenamefont {Blum}\ \emph {et~al.}(2023)\citenamefont {Blum} \emph
  {et~al.}}]{RBC:2023pvn}%
  \BibitemOpen
  \bibfield  {author} {\bibinfo {author} {\bibfnamefont {T.}~\bibnamefont
  {Blum}} \emph {et~al.} (\bibinfo {collaboration} {RBC, UKQCD}),\ }\href
  {https://doi.org/10.1103/PhysRevD.108.054507} {\bibfield  {journal} {\bibinfo
   {journal} {Phys. Rev. D}\ }\textbf {\bibinfo {volume} {108}},\ \bibinfo
  {pages} {054507} (\bibinfo {year} {2023})},\ \Eprint
  {https://arxiv.org/abs/2301.08696} {arXiv:2301.08696 [hep-lat]} \BibitemShut
  {NoStop}%
\bibitem [{\citenamefont {Kuberski}\ \emph {et~al.}(2024)\citenamefont
  {Kuberski}, \citenamefont {C\`e}, \citenamefont {von Hippel}, \citenamefont
  {Meyer}, \citenamefont {Ottnad}, \citenamefont {Risch},\ and\ \citenamefont
  {Wittig}}]{Kuberski:2024bcj}%
  \BibitemOpen
  \bibfield  {author} {\bibinfo {author} {\bibfnamefont {S.}~\bibnamefont
  {Kuberski}}, \bibinfo {author} {\bibfnamefont {M.}~\bibnamefont {C\`e}},
  \bibinfo {author} {\bibfnamefont {G.}~\bibnamefont {von Hippel}}, \bibinfo
  {author} {\bibfnamefont {H.~B.}\ \bibnamefont {Meyer}}, \bibinfo {author}
  {\bibfnamefont {K.}~\bibnamefont {Ottnad}}, \bibinfo {author} {\bibfnamefont
  {A.}~\bibnamefont {Risch}},\ and\ \bibinfo {author} {\bibfnamefont
  {H.}~\bibnamefont {Wittig}},\ }\href
  {https://doi.org/10.1007/JHEP03(2024)172} {\bibfield  {journal} {\bibinfo
  {journal} {JHEP}\ }\textbf {\bibinfo {volume} {03}},\ \bibinfo {pages}
  {172}},\ \Eprint {https://arxiv.org/abs/2401.11895} {arXiv:2401.11895
  [hep-lat]} \BibitemShut {NoStop}%
\bibitem [{\citenamefont {Davies}\ \emph {et~al.}(2022)\citenamefont {Davies}
  \emph {et~al.}}]{FermilabLattice:2022izv}%
  \BibitemOpen
  \bibfield  {author} {\bibinfo {author} {\bibfnamefont {C.~T.~H.}\
  \bibnamefont {Davies}} \emph {et~al.} (\bibinfo {collaboration} {Fermilab
  Lattice, MILC, HPQCD}),\ }\href {https://doi.org/10.1103/PhysRevD.106.074509}
  {\bibfield  {journal} {\bibinfo  {journal} {Phys. Rev. D}\ }\textbf {\bibinfo
  {volume} {106}},\ \bibinfo {pages} {074509} (\bibinfo {year} {2022})},\
  \Eprint {https://arxiv.org/abs/2207.04765} {arXiv:2207.04765 [hep-lat]}
  \BibitemShut {NoStop}%
\bibitem [{\citenamefont {Colangelo}\ \emph {et~al.}(2022)\citenamefont
  {Colangelo}, \citenamefont {El-Khadra}, \citenamefont {Hoferichter},
  \citenamefont {Keshavarzi}, \citenamefont {Lehner}, \citenamefont {Stoffer},\
  and\ \citenamefont {Teubner}}]{Colangelo:2022vok}%
  \BibitemOpen
  \bibfield  {author} {\bibinfo {author} {\bibfnamefont {G.}~\bibnamefont
  {Colangelo}}, \bibinfo {author} {\bibfnamefont {A.~X.}\ \bibnamefont
  {El-Khadra}}, \bibinfo {author} {\bibfnamefont {M.}~\bibnamefont
  {Hoferichter}}, \bibinfo {author} {\bibfnamefont {A.}~\bibnamefont
  {Keshavarzi}}, \bibinfo {author} {\bibfnamefont {C.}~\bibnamefont {Lehner}},
  \bibinfo {author} {\bibfnamefont {P.}~\bibnamefont {Stoffer}},\ and\ \bibinfo
  {author} {\bibfnamefont {T.}~\bibnamefont {Teubner}},\ }\href
  {https://doi.org/10.1016/j.physletb.2022.137313} {\bibfield  {journal}
  {\bibinfo  {journal} {Phys. Lett. B}\ }\textbf {\bibinfo {volume} {833}},\
  \bibinfo {pages} {137313} (\bibinfo {year} {2022})},\ \Eprint
  {https://arxiv.org/abs/2205.12963} {arXiv:2205.12963 [hep-ph]} \BibitemShut
  {NoStop}%
\bibitem [{\citenamefont {Blum}\ \emph {et~al.}(2024)\citenamefont {Blum} \emph
  {et~al.}}]{RBC:2024fic}%
  \BibitemOpen
  \bibfield  {author} {\bibinfo {author} {\bibfnamefont {T.}~\bibnamefont
  {Blum}} \emph {et~al.} (\bibinfo {collaboration} {RBC, UKQCD}),\ }\href@noop
  {} {\  (\bibinfo {year} {2024})},\ \Eprint {https://arxiv.org/abs/2410.20590}
  {arXiv:2410.20590 [hep-lat]} \BibitemShut {NoStop}%
\bibitem [{\citenamefont {Djukanovic}\ \emph {et~al.}(2024)\citenamefont
  {Djukanovic}, \citenamefont {von Hippel}, \citenamefont {Kuberski},
  \citenamefont {Meyer}, \citenamefont {Miller}, \citenamefont {Ottnad},
  \citenamefont {Parrino}, \citenamefont {Risch},\ and\ \citenamefont
  {Wittig}}]{Djukanovic:2024cmq}%
  \BibitemOpen
  \bibfield  {author} {\bibinfo {author} {\bibfnamefont {D.}~\bibnamefont
  {Djukanovic}}, \bibinfo {author} {\bibfnamefont {G.}~\bibnamefont {von
  Hippel}}, \bibinfo {author} {\bibfnamefont {S.}~\bibnamefont {Kuberski}},
  \bibinfo {author} {\bibfnamefont {H.~B.}\ \bibnamefont {Meyer}}, \bibinfo
  {author} {\bibfnamefont {N.}~\bibnamefont {Miller}}, \bibinfo {author}
  {\bibfnamefont {K.}~\bibnamefont {Ottnad}}, \bibinfo {author} {\bibfnamefont
  {J.}~\bibnamefont {Parrino}}, \bibinfo {author} {\bibfnamefont
  {A.}~\bibnamefont {Risch}},\ and\ \bibinfo {author} {\bibfnamefont
  {H.}~\bibnamefont {Wittig}},\ }\href@noop {} {\  (\bibinfo {year} {2024})},\
  \Eprint {https://arxiv.org/abs/2411.07969} {arXiv:2411.07969 [hep-lat]}
  \BibitemShut {NoStop}%
\bibitem [{\citenamefont {Passera}\ \emph {et~al.}(2008)\citenamefont
  {Passera}, \citenamefont {Marciano},\ and\ \citenamefont
  {Sirlin}}]{Passera:2008jk}%
  \BibitemOpen
  \bibfield  {author} {\bibinfo {author} {\bibfnamefont {M.}~\bibnamefont
  {Passera}}, \bibinfo {author} {\bibfnamefont {W.~J.}\ \bibnamefont
  {Marciano}},\ and\ \bibinfo {author} {\bibfnamefont {A.}~\bibnamefont
  {Sirlin}},\ }\href {https://doi.org/10.1103/PhysRevD.78.013009} {\bibfield
  {journal} {\bibinfo  {journal} {Phys. Rev. D}\ }\textbf {\bibinfo {volume}
  {78}},\ \bibinfo {pages} {013009} (\bibinfo {year} {2008})},\ \Eprint
  {https://arxiv.org/abs/0804.1142} {arXiv:0804.1142 [hep-ph]} \BibitemShut
  {NoStop}%
\bibitem [{\citenamefont {Crivellin}\ \emph {et~al.}(2020)\citenamefont
  {Crivellin}, \citenamefont {Hoferichter}, \citenamefont {Manzari},\ and\
  \citenamefont {Montull}}]{Crivellin:2020zul}%
  \BibitemOpen
  \bibfield  {author} {\bibinfo {author} {\bibfnamefont {A.}~\bibnamefont
  {Crivellin}}, \bibinfo {author} {\bibfnamefont {M.}~\bibnamefont
  {Hoferichter}}, \bibinfo {author} {\bibfnamefont {C.~A.}\ \bibnamefont
  {Manzari}},\ and\ \bibinfo {author} {\bibfnamefont {M.}~\bibnamefont
  {Montull}},\ }\href {https://doi.org/10.1103/PhysRevLett.125.091801}
  {\bibfield  {journal} {\bibinfo  {journal} {Phys. Rev. Lett.}\ }\textbf
  {\bibinfo {volume} {125}},\ \bibinfo {pages} {091801} (\bibinfo {year}
  {2020})},\ \Eprint {https://arxiv.org/abs/2003.04886} {arXiv:2003.04886
  [hep-ph]} \BibitemShut {NoStop}%
\bibitem [{\citenamefont {Keshavarzi}\ \emph
  {et~al.}(2020{\natexlab{b}})\citenamefont {Keshavarzi}, \citenamefont
  {Marciano}, \citenamefont {Passera},\ and\ \citenamefont
  {Sirlin}}]{Keshavarzi:2020bfy}%
  \BibitemOpen
  \bibfield  {author} {\bibinfo {author} {\bibfnamefont {A.}~\bibnamefont
  {Keshavarzi}}, \bibinfo {author} {\bibfnamefont {W.~J.}\ \bibnamefont
  {Marciano}}, \bibinfo {author} {\bibfnamefont {M.}~\bibnamefont {Passera}},\
  and\ \bibinfo {author} {\bibfnamefont {A.}~\bibnamefont {Sirlin}},\ }\href
  {https://doi.org/10.1103/PhysRevD.102.033002} {\bibfield  {journal} {\bibinfo
   {journal} {Phys. Rev. D}\ }\textbf {\bibinfo {volume} {102}},\ \bibinfo
  {pages} {033002} (\bibinfo {year} {2020}{\natexlab{b}})},\ \Eprint
  {https://arxiv.org/abs/2006.12666} {arXiv:2006.12666 [hep-ph]} \BibitemShut
  {NoStop}%
\bibitem [{\citenamefont {Malaescu}\ and\ \citenamefont
  {Schott}(2021)}]{Malaescu:2020zuc}%
  \BibitemOpen
  \bibfield  {author} {\bibinfo {author} {\bibfnamefont {B.}~\bibnamefont
  {Malaescu}}\ and\ \bibinfo {author} {\bibfnamefont {M.}~\bibnamefont
  {Schott}},\ }\href {https://doi.org/10.1140/epjc/s10052-021-08848-9}
  {\bibfield  {journal} {\bibinfo  {journal} {Eur. Phys. J. C}\ }\textbf
  {\bibinfo {volume} {81}},\ \bibinfo {pages} {46} (\bibinfo {year} {2021})},\
  \Eprint {https://arxiv.org/abs/2008.08107} {arXiv:2008.08107 [hep-ph]}
  \BibitemShut {NoStop}%
\bibitem [{\citenamefont {Colangelo}\ \emph {et~al.}(2021)\citenamefont
  {Colangelo}, \citenamefont {Hoferichter},\ and\ \citenamefont
  {Stoffer}}]{Colangelo:2020lcg}%
  \BibitemOpen
  \bibfield  {author} {\bibinfo {author} {\bibfnamefont {G.}~\bibnamefont
  {Colangelo}}, \bibinfo {author} {\bibfnamefont {M.}~\bibnamefont
  {Hoferichter}},\ and\ \bibinfo {author} {\bibfnamefont {P.}~\bibnamefont
  {Stoffer}},\ }\href {https://doi.org/10.1016/j.physletb.2021.136073}
  {\bibfield  {journal} {\bibinfo  {journal} {Phys. Lett. B}\ }\textbf
  {\bibinfo {volume} {814}},\ \bibinfo {pages} {136073} (\bibinfo {year}
  {2021})},\ \Eprint {https://arxiv.org/abs/2010.07943} {arXiv:2010.07943
  [hep-ph]} \BibitemShut {NoStop}%
\bibitem [{\citenamefont {Benton}\ \emph {et~al.}(2023)\citenamefont {Benton},
  \citenamefont {Boito}, \citenamefont {Golterman}, \citenamefont {Keshavarzi},
  \citenamefont {Maltman},\ and\ \citenamefont {Peris}}]{Benton:2023dci}%
  \BibitemOpen
  \bibfield  {author} {\bibinfo {author} {\bibfnamefont {G.}~\bibnamefont
  {Benton}}, \bibinfo {author} {\bibfnamefont {D.}~\bibnamefont {Boito}},
  \bibinfo {author} {\bibfnamefont {M.}~\bibnamefont {Golterman}}, \bibinfo
  {author} {\bibfnamefont {A.}~\bibnamefont {Keshavarzi}}, \bibinfo {author}
  {\bibfnamefont {K.}~\bibnamefont {Maltman}},\ and\ \bibinfo {author}
  {\bibfnamefont {S.}~\bibnamefont {Peris}},\ }\href
  {https://doi.org/10.1103/PhysRevLett.131.251803} {\bibfield  {journal}
  {\bibinfo  {journal} {Phys. Rev. Lett.}\ }\textbf {\bibinfo {volume} {131}},\
  \bibinfo {pages} {251803} (\bibinfo {year} {2023})},\ \Eprint
  {https://arxiv.org/abs/2306.16808} {arXiv:2306.16808 [hep-ph]} \BibitemShut
  {NoStop}%
\bibitem [{\citenamefont {Benton}\ \emph {et~al.}(2024)\citenamefont {Benton},
  \citenamefont {Boito}, \citenamefont {Golterman}, \citenamefont {Keshavarzi},
  \citenamefont {Maltman},\ and\ \citenamefont {Peris}}]{Benton:2023fcv}%
  \BibitemOpen
  \bibfield  {author} {\bibinfo {author} {\bibfnamefont {G.}~\bibnamefont
  {Benton}}, \bibinfo {author} {\bibfnamefont {D.}~\bibnamefont {Boito}},
  \bibinfo {author} {\bibfnamefont {M.}~\bibnamefont {Golterman}}, \bibinfo
  {author} {\bibfnamefont {A.}~\bibnamefont {Keshavarzi}}, \bibinfo {author}
  {\bibfnamefont {K.}~\bibnamefont {Maltman}},\ and\ \bibinfo {author}
  {\bibfnamefont {S.}~\bibnamefont {Peris}},\ }\href
  {https://doi.org/10.1103/PhysRevD.109.036010} {\bibfield  {journal} {\bibinfo
   {journal} {Phys. Rev. D}\ }\textbf {\bibinfo {volume} {109}},\ \bibinfo
  {pages} {036010} (\bibinfo {year} {2024})},\ \Eprint
  {https://arxiv.org/abs/2311.09523} {arXiv:2311.09523 [hep-ph]} \BibitemShut
  {NoStop}%
\bibitem [{\citenamefont {Di~Luzio}\ \emph {et~al.}(2022)\citenamefont
  {Di~Luzio}, \citenamefont {Masiero}, \citenamefont {Paradisi},\ and\
  \citenamefont {Passera}}]{DiLuzio:2021uty}%
  \BibitemOpen
  \bibfield  {author} {\bibinfo {author} {\bibfnamefont {L.}~\bibnamefont
  {Di~Luzio}}, \bibinfo {author} {\bibfnamefont {A.}~\bibnamefont {Masiero}},
  \bibinfo {author} {\bibfnamefont {P.}~\bibnamefont {Paradisi}},\ and\
  \bibinfo {author} {\bibfnamefont {M.}~\bibnamefont {Passera}},\ }\href
  {https://doi.org/10.1016/j.physletb.2022.137037} {\bibfield  {journal}
  {\bibinfo  {journal} {Phys. Lett. B}\ }\textbf {\bibinfo {volume} {829}},\
  \bibinfo {pages} {137037} (\bibinfo {year} {2022})},\ \Eprint
  {https://arxiv.org/abs/2112.08312} {arXiv:2112.08312 [hep-ph]} \BibitemShut
  {NoStop}%
\bibitem [{\citenamefont {Darm\'e}\ \emph {et~al.}(2022)\citenamefont
  {Darm\'e}, \citenamefont {Grilli~di Cortona},\ and\ \citenamefont
  {Nardi}}]{Darme:2021huc}%
  \BibitemOpen
  \bibfield  {author} {\bibinfo {author} {\bibfnamefont {L.}~\bibnamefont
  {Darm\'e}}, \bibinfo {author} {\bibfnamefont {G.}~\bibnamefont {Grilli~di
  Cortona}},\ and\ \bibinfo {author} {\bibfnamefont {E.}~\bibnamefont
  {Nardi}},\ }\href {https://doi.org/10.1007/JHEP06(2022)122} {\bibfield
  {journal} {\bibinfo  {journal} {JHEP}\ }\textbf {\bibinfo {volume} {06}},\
  \bibinfo {pages} {122}},\ \Eprint {https://arxiv.org/abs/2112.09139}
  {arXiv:2112.09139 [hep-ph]} \BibitemShut {NoStop}%
\bibitem [{\citenamefont {Crivellin}\ and\ \citenamefont
  {Hoferichter}(2023)}]{Crivellin:2022gfu}%
  \BibitemOpen
  \bibfield  {author} {\bibinfo {author} {\bibfnamefont {A.}~\bibnamefont
  {Crivellin}}\ and\ \bibinfo {author} {\bibfnamefont {M.}~\bibnamefont
  {Hoferichter}},\ }\href {https://doi.org/10.1103/PhysRevD.108.013005}
  {\bibfield  {journal} {\bibinfo  {journal} {Phys. Rev. D}\ }\textbf {\bibinfo
  {volume} {108}},\ \bibinfo {pages} {013005} (\bibinfo {year} {2023})},\
  \Eprint {https://arxiv.org/abs/2211.12516} {arXiv:2211.12516 [hep-ph]}
  \BibitemShut {NoStop}%
\bibitem [{\citenamefont {Darm\'e}\ \emph {et~al.}(2023)\citenamefont
  {Darm\'e}, \citenamefont {Grilli~di Cortona},\ and\ \citenamefont
  {Nardi}}]{Darme:2022yal}%
  \BibitemOpen
  \bibfield  {author} {\bibinfo {author} {\bibfnamefont {L.}~\bibnamefont
  {Darm\'e}}, \bibinfo {author} {\bibfnamefont {G.}~\bibnamefont {Grilli~di
  Cortona}},\ and\ \bibinfo {author} {\bibfnamefont {E.}~\bibnamefont
  {Nardi}},\ }\href {https://doi.org/10.1103/PhysRevD.108.095056} {\bibfield
  {journal} {\bibinfo  {journal} {Phys. Rev. D}\ }\textbf {\bibinfo {volume}
  {108}},\ \bibinfo {pages} {095056} (\bibinfo {year} {2023})},\ \Eprint
  {https://arxiv.org/abs/2212.03877} {arXiv:2212.03877 [hep-ph]} \BibitemShut
  {NoStop}%
\bibitem [{\citenamefont {Coyle}\ and\ \citenamefont
  {Wagner}(2023)}]{Coyle:2023nmi}%
  \BibitemOpen
  \bibfield  {author} {\bibinfo {author} {\bibfnamefont {N.~M.}\ \bibnamefont
  {Coyle}}\ and\ \bibinfo {author} {\bibfnamefont {C.~E.~M.}\ \bibnamefont
  {Wagner}},\ }\href {https://doi.org/10.1007/JHEP12(2023)071} {\bibfield
  {journal} {\bibinfo  {journal} {JHEP}\ }\textbf {\bibinfo {volume} {12}},\
  \bibinfo {pages} {071}},\ \Eprint {https://arxiv.org/abs/2305.02354}
  {arXiv:2305.02354 [hep-ph]} \BibitemShut {NoStop}%
\bibitem [{\citenamefont {Ignatov}\ \emph
  {et~al.}(2024{\natexlab{a}})\citenamefont {Ignatov} \emph
  {et~al.}}]{CMD-3:2023alj}%
  \BibitemOpen
  \bibfield  {author} {\bibinfo {author} {\bibfnamefont {F.~V.}\ \bibnamefont
  {Ignatov}} \emph {et~al.} (\bibinfo {collaboration} {CMD-3}),\ }\href
  {https://doi.org/10.1103/PhysRevD.109.112002} {\bibfield  {journal} {\bibinfo
   {journal} {Phys. Rev. D}\ }\textbf {\bibinfo {volume} {109}},\ \bibinfo
  {pages} {112002} (\bibinfo {year} {2024}{\natexlab{a}})},\ \Eprint
  {https://arxiv.org/abs/2302.08834} {arXiv:2302.08834 [hep-ex]} \BibitemShut
  {NoStop}%
\bibitem [{\citenamefont {Ignatov}\ \emph
  {et~al.}(2024{\natexlab{b}})\citenamefont {Ignatov} \emph
  {et~al.}}]{CMD-3:2023rfe}%
  \BibitemOpen
  \bibfield  {author} {\bibinfo {author} {\bibfnamefont {F.~V.}\ \bibnamefont
  {Ignatov}} \emph {et~al.} (\bibinfo {collaboration} {CMD-3}),\ }\href
  {https://doi.org/10.1103/PhysRevLett.132.231903} {\bibfield  {journal}
  {\bibinfo  {journal} {Phys. Rev. Lett.}\ }\textbf {\bibinfo {volume} {132}},\
  \bibinfo {pages} {231903} (\bibinfo {year} {2024}{\natexlab{b}})},\ \Eprint
  {https://arxiv.org/abs/2309.12910} {arXiv:2309.12910 [hep-ex]} \BibitemShut
  {NoStop}%
\bibitem [{\citenamefont {Achasov}\ \emph {et~al.}(2006)\citenamefont {Achasov}
  \emph {et~al.}}]{Achasov:2006vp}%
  \BibitemOpen
  \bibfield  {author} {\bibinfo {author} {\bibfnamefont {M.~N.}\ \bibnamefont
  {Achasov}} \emph {et~al.},\ }\href
  {https://doi.org/10.1134/S106377610609007X} {\bibfield  {journal} {\bibinfo
  {journal} {J. Exp. Theor. Phys.}\ }\textbf {\bibinfo {volume} {103}},\
  \bibinfo {pages} {380} (\bibinfo {year} {2006})},\ \Eprint
  {https://arxiv.org/abs/hep-ex/0605013} {arXiv:hep-ex/0605013} \BibitemShut
  {NoStop}%
\bibitem [{\citenamefont {Aul'chenko}\ \emph {et~al.}(2006)\citenamefont
  {Aul'chenko} \emph {et~al.}}]{Aulchenko:2006dxz}%
  \BibitemOpen
  \bibfield  {author} {\bibinfo {author} {\bibfnamefont {V.~M.}\ \bibnamefont
  {Aul'chenko}} \emph {et~al.},\ }\href
  {https://doi.org/10.1134/S0021364006200021} {\bibfield  {journal} {\bibinfo
  {journal} {JETP Lett.}\ }\textbf {\bibinfo {volume} {84}},\ \bibinfo {pages}
  {413} (\bibinfo {year} {2006})},\ \Eprint
  {https://arxiv.org/abs/hep-ex/0610016} {arXiv:hep-ex/0610016} \BibitemShut
  {NoStop}%
\bibitem [{\citenamefont {Akhmetshin}\ \emph {et~al.}(2007)\citenamefont
  {Akhmetshin} \emph {et~al.}}]{CMD-2:2006gxt}%
  \BibitemOpen
  \bibfield  {author} {\bibinfo {author} {\bibfnamefont {R.~R.}\ \bibnamefont
  {Akhmetshin}} \emph {et~al.} (\bibinfo {collaboration} {CMD-2}),\ }\href
  {https://doi.org/10.1016/j.physletb.2007.01.073} {\bibfield  {journal}
  {\bibinfo  {journal} {Phys. Lett. B}\ }\textbf {\bibinfo {volume} {648}},\
  \bibinfo {pages} {28} (\bibinfo {year} {2007})},\ \Eprint
  {https://arxiv.org/abs/hep-ex/0610021} {arXiv:hep-ex/0610021} \BibitemShut
  {NoStop}%
\bibitem [{\citenamefont {Ambrosino}\ \emph {et~al.}(2009)\citenamefont
  {Ambrosino} \emph {et~al.}}]{KLOE:2008fmq}%
  \BibitemOpen
  \bibfield  {author} {\bibinfo {author} {\bibfnamefont {F.}~\bibnamefont
  {Ambrosino}} \emph {et~al.} (\bibinfo {collaboration} {KLOE}),\ }\href
  {https://doi.org/10.1016/j.physletb.2008.10.060} {\bibfield  {journal}
  {\bibinfo  {journal} {Phys. Lett. B}\ }\textbf {\bibinfo {volume} {670}},\
  \bibinfo {pages} {285} (\bibinfo {year} {2009})},\ \Eprint
  {https://arxiv.org/abs/0809.3950} {arXiv:0809.3950 [hep-ex]} \BibitemShut
  {NoStop}%
\bibitem [{\citenamefont {Ambrosino}\ \emph {et~al.}(2011)\citenamefont
  {Ambrosino} \emph {et~al.}}]{KLOE:2010qei}%
  \BibitemOpen
  \bibfield  {author} {\bibinfo {author} {\bibfnamefont {F.}~\bibnamefont
  {Ambrosino}} \emph {et~al.} (\bibinfo {collaboration} {KLOE}),\ }\href
  {https://doi.org/10.1016/j.physletb.2011.04.055} {\bibfield  {journal}
  {\bibinfo  {journal} {Phys. Lett. B}\ }\textbf {\bibinfo {volume} {700}},\
  \bibinfo {pages} {102} (\bibinfo {year} {2011})},\ \Eprint
  {https://arxiv.org/abs/1006.5313} {arXiv:1006.5313 [hep-ex]} \BibitemShut
  {NoStop}%
\bibitem [{\citenamefont {Babusci}\ \emph {et~al.}(2013)\citenamefont {Babusci}
  \emph {et~al.}}]{KLOE:2012anl}%
  \BibitemOpen
  \bibfield  {author} {\bibinfo {author} {\bibfnamefont {D.}~\bibnamefont
  {Babusci}} \emph {et~al.} (\bibinfo {collaboration} {KLOE}),\ }\href
  {https://doi.org/10.1016/j.physletb.2013.02.029} {\bibfield  {journal}
  {\bibinfo  {journal} {Phys. Lett. B}\ }\textbf {\bibinfo {volume} {720}},\
  \bibinfo {pages} {336} (\bibinfo {year} {2013})},\ \Eprint
  {https://arxiv.org/abs/1212.4524} {arXiv:1212.4524 [hep-ex]} \BibitemShut
  {NoStop}%
\bibitem [{\citenamefont {Anastasi}\ \emph {et~al.}(2018)\citenamefont
  {Anastasi} \emph {et~al.}}]{KLOE-2:2017fda}%
  \BibitemOpen
  \bibfield  {author} {\bibinfo {author} {\bibfnamefont {A.}~\bibnamefont
  {Anastasi}} \emph {et~al.} (\bibinfo {collaboration} {KLOE-2}),\ }\href
  {https://doi.org/10.1007/JHEP03(2018)173} {\bibfield  {journal} {\bibinfo
  {journal} {JHEP}\ }\textbf {\bibinfo {volume} {03}},\ \bibinfo {pages}
  {173}},\ \Eprint {https://arxiv.org/abs/1711.03085} {arXiv:1711.03085
  [hep-ex]} \BibitemShut {NoStop}%
\bibitem [{\citenamefont {Aubert}\ \emph {et~al.}(2009)\citenamefont {Aubert}
  \emph {et~al.}}]{BaBar:2009wpw}%
  \BibitemOpen
  \bibfield  {author} {\bibinfo {author} {\bibfnamefont {B.}~\bibnamefont
  {Aubert}} \emph {et~al.} (\bibinfo {collaboration} {BaBar}),\ }\href
  {https://doi.org/10.1103/PhysRevLett.103.231801} {\bibfield  {journal}
  {\bibinfo  {journal} {Phys. Rev. Lett.}\ }\textbf {\bibinfo {volume} {103}},\
  \bibinfo {pages} {231801} (\bibinfo {year} {2009})},\ \Eprint
  {https://arxiv.org/abs/0908.3589} {arXiv:0908.3589 [hep-ex]} \BibitemShut
  {NoStop}%
\bibitem [{\citenamefont {Lees}\ \emph {et~al.}(2012)\citenamefont {Lees} \emph
  {et~al.}}]{BaBar:2012bdw}%
  \BibitemOpen
  \bibfield  {author} {\bibinfo {author} {\bibfnamefont {J.~P.}\ \bibnamefont
  {Lees}} \emph {et~al.} (\bibinfo {collaboration} {BaBar}),\ }\href
  {https://doi.org/10.1103/PhysRevD.86.032013} {\bibfield  {journal} {\bibinfo
  {journal} {Phys. Rev. D}\ }\textbf {\bibinfo {volume} {86}},\ \bibinfo
  {pages} {032013} (\bibinfo {year} {2012})},\ \Eprint
  {https://arxiv.org/abs/1205.2228} {arXiv:1205.2228 [hep-ex]} \BibitemShut
  {NoStop}%
\bibitem [{\citenamefont {Ablikim}\ \emph {et~al.}(2016)\citenamefont {Ablikim}
  \emph {et~al.}}]{BESIII:2015equ}%
  \BibitemOpen
  \bibfield  {author} {\bibinfo {author} {\bibfnamefont {M.}~\bibnamefont
  {Ablikim}} \emph {et~al.} (\bibinfo {collaboration} {BESIII}),\ }\href
  {https://doi.org/10.1016/j.physletb.2015.11.043} {\bibfield  {journal}
  {\bibinfo  {journal} {Phys. Lett. B}\ }\textbf {\bibinfo {volume} {753}},\
  \bibinfo {pages} {629} (\bibinfo {year} {2016})},\ \bibinfo {note} {[Erratum:
  Phys.Lett.B 812, 135982 (2021)]},\ \Eprint {https://arxiv.org/abs/1507.08188}
  {arXiv:1507.08188 [hep-ex]} \BibitemShut {NoStop}%
\bibitem [{\citenamefont {Xiao}\ \emph {et~al.}(2018)\citenamefont {Xiao},
  \citenamefont {Dobbs}, \citenamefont {Tomaradze}, \citenamefont {Seth},\ and\
  \citenamefont {Bonvicini}}]{Xiao:2017dqv}%
  \BibitemOpen
  \bibfield  {author} {\bibinfo {author} {\bibfnamefont {T.}~\bibnamefont
  {Xiao}}, \bibinfo {author} {\bibfnamefont {S.}~\bibnamefont {Dobbs}},
  \bibinfo {author} {\bibfnamefont {A.}~\bibnamefont {Tomaradze}}, \bibinfo
  {author} {\bibfnamefont {K.~K.}\ \bibnamefont {Seth}},\ and\ \bibinfo
  {author} {\bibfnamefont {G.}~\bibnamefont {Bonvicini}},\ }\href
  {https://doi.org/10.1103/PhysRevD.97.032012} {\bibfield  {journal} {\bibinfo
  {journal} {Phys. Rev. D}\ }\textbf {\bibinfo {volume} {97}},\ \bibinfo
  {pages} {032012} (\bibinfo {year} {2018})},\ \Eprint
  {https://arxiv.org/abs/1712.04530} {arXiv:1712.04530 [hep-ex]} \BibitemShut
  {NoStop}%
\bibitem [{\citenamefont {Achasov}\ \emph {et~al.}(2021)\citenamefont {Achasov}
  \emph {et~al.}}]{SND:2020nwa}%
  \BibitemOpen
  \bibfield  {author} {\bibinfo {author} {\bibfnamefont {M.~N.}\ \bibnamefont
  {Achasov}} \emph {et~al.} (\bibinfo {collaboration} {SND}),\ }\href
  {https://doi.org/10.1007/JHEP01(2021)113} {\bibfield  {journal} {\bibinfo
  {journal} {JHEP}\ }\textbf {\bibinfo {volume} {01}},\ \bibinfo {pages}
  {113}},\ \Eprint {https://arxiv.org/abs/2004.00263} {arXiv:2004.00263
  [hep-ex]} \BibitemShut {NoStop}%
\bibitem [{\citenamefont {{Zhiqing ZHANG}}(2024)}]{newBaBar}%
  \BibitemOpen
  \bibfield  {author} {\bibinfo {author} {\bibnamefont {{Zhiqing ZHANG}}},\
  }\href@noop {} {\bibinfo {title} {News from {BABAR}}},\ \bibinfo
  {howpublished} {{M}uon $g$$-$$2$ {T}heory {I}nitiative {S}pring 2024
  {M}eeting,
  \textsc{url:}~\url{https://indico.cern.ch/event/1400808/contributions/5902094/attachments/2842140/4968393/Zhang_HVP240422.pdf}}
  (\bibinfo {year} {2024})\BibitemShut {NoStop}%
\bibitem [{\citenamefont {{Hisaki Hayashii}}(2024)}]{newBelleII}%
  \BibitemOpen
  \bibfield  {author} {\bibinfo {author} {\bibnamefont {{Hisaki Hayashii}}},\
  }\href@noop {} {\bibinfo {title} {Status and plans regarding $g$$-$$2$ at
  belle {II}}},\ \bibinfo {howpublished} {{M}uon $g$$-$$2$ {T}heory
  {I}nitiative {S}pring 2024 {M}eeting,
  \textsc{url:}~\url{https://indico.cern.ch/event/1400808/contributions/5902135/attachments/2841580/4968265/g-2-theory_mni-workshop_240422_v2.pdf}}
  (\bibinfo {year} {2024})\BibitemShut {NoStop}%
\bibitem [{\citenamefont {{Riccardo Aliberti}}(2024)}]{newBESIII}%
  \BibitemOpen
  \bibfield  {author} {\bibinfo {author} {\bibnamefont {{Riccardo Aliberti}}},\
  }\href@noop {} {\bibinfo {title} {{S}tatus and {P}lans for {E}xperimental
  {I}nputs to {HVP} at {BESIII}}},\ \bibinfo {howpublished} {{M}uon $g$$-$$2$
  {T}heory {I}nitiative {S}pring 2024 {M}eeting,
  \textsc{url:}~\url{https://indico.cern.ch/event/1400808/contributions/5902095/attachments/2841881/4967869/Muon_g-2_spring_meeting_2024.pdf}}
  (\bibinfo {year} {2024})\BibitemShut {NoStop}%
\bibitem [{\citenamefont {{Ivan Logashenko}}(2024)}]{newCMD3}%
  \BibitemOpen
  \bibfield  {author} {\bibinfo {author} {\bibnamefont {{Ivan Logashenko}}},\
  }\href@noop {} {\bibinfo {title} {{CMD2/3} report (on $\pi^+\pi^-$)}},\
  \bibinfo {howpublished} {{M}uon $g$$-$$2$ {T}heory {I}nitiative {S}pring 2024
  {M}eeting,
  \textsc{url:}~\url{https://indico.cern.ch/event/1400808/contributions/5902096/attachments/2842193/4968485/Logashenko_CMD_TI_2024.pdf}}
  (\bibinfo {year} {2024})\BibitemShut {NoStop}%
\bibitem [{\citenamefont {{Giuseppe Mandaglio}}(2023)}]{newKLOE}%
  \BibitemOpen
  \bibfield  {author} {\bibinfo {author} {\bibnamefont {{Giuseppe
  Mandaglio}}},\ }\href@noop {} {\bibinfo {title} {Hadron physics results at
  {KLOE-2}}},\ \bibinfo {howpublished} {{S}ixth Plenary Workshop of the {M}uon
  $g$$-$$2$ {T}heory {I}nitiative (2023),
  \textsc{url:}~\url{https://indico.cern.ch/event/1400808/contributions/5902134/attachments/2841711/4967471/talk_IL.pdf}}
  (\bibinfo {year} {2023})\BibitemShut {NoStop}%
\bibitem [{\citenamefont {{Andrey Kupich}}(2024)}]{newSND}%
  \BibitemOpen
  \bibfield  {author} {\bibinfo {author} {\bibnamefont {{Andrey Kupich}}},\
  }\href@noop {} {\bibinfo {title} {Preliminary results of the $e^+e^- \to
  \pi^+\pi^-$ analysis with {SND} at {VEPP-2000}}},\ \bibinfo {howpublished}
  {{M}uon $g$$-$$2$ {T}heory {I}nitiative {S}pring 2024 {M}eeting,
  \textsc{url:}~\url{https://indico.cern.ch/event/1400808/contributions/5902134/attachments/2841711/4967471/talk_IL.pdf}}
  (\bibinfo {year} {2024})\BibitemShut {NoStop}%
\bibitem [{\citenamefont {Lees}\ \emph {et~al.}(2023)\citenamefont {Lees} \emph
  {et~al.}}]{BaBar:2023xiy}%
  \BibitemOpen
  \bibfield  {author} {\bibinfo {author} {\bibfnamefont {J.~P.}\ \bibnamefont
  {Lees}} \emph {et~al.} (\bibinfo {collaboration} {BaBar}),\ }\href
  {https://doi.org/10.1103/PhysRevD.108.L111103} {\bibfield  {journal}
  {\bibinfo  {journal} {Phys. Rev. D}\ }\textbf {\bibinfo {volume} {108}},\
  \bibinfo {pages} {L111103} (\bibinfo {year} {2023})},\ \Eprint
  {https://arxiv.org/abs/2308.05233} {arXiv:2308.05233 [hep-ex]} \BibitemShut
  {NoStop}%
\bibitem [{\citenamefont {Davier}\ \emph {et~al.}(2024)\citenamefont {Davier},
  \citenamefont {Hoecker}, \citenamefont {Lutz}, \citenamefont {Malaescu},\
  and\ \citenamefont {Zhang}}]{Davier:2023fpl}%
  \BibitemOpen
  \bibfield  {author} {\bibinfo {author} {\bibfnamefont {M.}~\bibnamefont
  {Davier}}, \bibinfo {author} {\bibfnamefont {A.}~\bibnamefont {Hoecker}},
  \bibinfo {author} {\bibfnamefont {A.-M.}\ \bibnamefont {Lutz}}, \bibinfo
  {author} {\bibfnamefont {B.}~\bibnamefont {Malaescu}},\ and\ \bibinfo
  {author} {\bibfnamefont {Z.}~\bibnamefont {Zhang}},\ }\href
  {https://doi.org/10.1140/epjc/s10052-024-12964-7} {\bibfield  {journal}
  {\bibinfo  {journal} {Eur. Phys. J. C}\ }\textbf {\bibinfo {volume} {84}},\
  \bibinfo {pages} {721} (\bibinfo {year} {2024})},\ \Eprint
  {https://arxiv.org/abs/2312.02053} {arXiv:2312.02053 [hep-ph]} \BibitemShut
  {NoStop}%
\bibitem [{\citenamefont {Carloni~Calame}\ \emph {et~al.}(2015)\citenamefont
  {Carloni~Calame}, \citenamefont {Passera}, \citenamefont {Trentadue},\ and\
  \citenamefont {Venanzoni}}]{CarloniCalame:2015obs}%
  \BibitemOpen
  \bibfield  {author} {\bibinfo {author} {\bibfnamefont {C.~M.}\ \bibnamefont
  {Carloni~Calame}}, \bibinfo {author} {\bibfnamefont {M.}~\bibnamefont
  {Passera}}, \bibinfo {author} {\bibfnamefont {L.}~\bibnamefont {Trentadue}},\
  and\ \bibinfo {author} {\bibfnamefont {G.}~\bibnamefont {Venanzoni}},\ }\href
  {https://doi.org/10.1016/j.physletb.2015.05.020} {\bibfield  {journal}
  {\bibinfo  {journal} {Phys. Lett. B}\ }\textbf {\bibinfo {volume} {746}},\
  \bibinfo {pages} {325} (\bibinfo {year} {2015})},\ \Eprint
  {https://arxiv.org/abs/1504.02228} {arXiv:1504.02228 [hep-ph]} \BibitemShut
  {NoStop}%
\bibitem [{\citenamefont {Abbiendi}\ \emph {et~al.}(2017)\citenamefont
  {Abbiendi} \emph {et~al.}}]{Abbiendi:2016xup}%
  \BibitemOpen
  \bibfield  {author} {\bibinfo {author} {\bibfnamefont {G.}~\bibnamefont
  {Abbiendi}} \emph {et~al.},\ }\href
  {https://doi.org/10.1140/epjc/s10052-017-4633-z} {\bibfield  {journal}
  {\bibinfo  {journal} {Eur. Phys. J. C}\ }\textbf {\bibinfo {volume} {77}},\
  \bibinfo {pages} {139} (\bibinfo {year} {2017})},\ \Eprint
  {https://arxiv.org/abs/1609.08987} {arXiv:1609.08987 [hep-ex]} \BibitemShut
  {NoStop}%
\bibitem [{\citenamefont {Abbiendi}(2019)}]{Abbiendi:2677471}%
  \BibitemOpen
  \bibfield  {author} {\bibinfo {author} {\bibfnamefont {G.}~\bibnamefont
  {Abbiendi}},\ }\href {https://cds.cern.ch/record/2677471} {\emph {\bibinfo
  {title} {{Letter of Intent: the MUonE project}}}},\ \bibinfo {type} {Tech.
  Rep.}\ (\bibinfo  {institution} {CERN},\ \bibinfo {address} {Geneva},\
  \bibinfo {year} {2019})\BibitemShut {NoStop}%
\bibitem [{\citenamefont {Banerjee}\ \emph {et~al.}(2020)\citenamefont
  {Banerjee} \emph {et~al.}}]{Banerjee:2020tdt}%
  \BibitemOpen
  \bibfield  {author} {\bibinfo {author} {\bibfnamefont {P.}~\bibnamefont
  {Banerjee}} \emph {et~al.},\ }\href
  {https://doi.org/10.1140/epjc/s10052-020-8138-9} {\bibfield  {journal}
  {\bibinfo  {journal} {Eur. Phys. J. C}\ }\textbf {\bibinfo {volume} {80}},\
  \bibinfo {pages} {591} (\bibinfo {year} {2020})},\ \Eprint
  {https://arxiv.org/abs/2004.13663} {arXiv:2004.13663 [hep-ph]} \BibitemShut
  {NoStop}%
\bibitem [{\citenamefont {Masiero}\ \emph {et~al.}(2020)\citenamefont
  {Masiero}, \citenamefont {Paradisi},\ and\ \citenamefont
  {Passera}}]{Masiero:2020vxk}%
  \BibitemOpen
  \bibfield  {author} {\bibinfo {author} {\bibfnamefont {A.}~\bibnamefont
  {Masiero}}, \bibinfo {author} {\bibfnamefont {P.}~\bibnamefont {Paradisi}},\
  and\ \bibinfo {author} {\bibfnamefont {M.}~\bibnamefont {Passera}},\ }\href
  {https://doi.org/10.1103/PhysRevD.102.075013} {\bibfield  {journal} {\bibinfo
   {journal} {Phys. Rev. D}\ }\textbf {\bibinfo {volume} {102}},\ \bibinfo
  {pages} {075013} (\bibinfo {year} {2020})},\ \Eprint
  {https://arxiv.org/abs/2002.05418} {arXiv:2002.05418 [hep-ph]} \BibitemShut
  {NoStop}%
\bibitem [{\citenamefont {Masjuan}\ \emph {et~al.}(2024)\citenamefont
  {Masjuan}, \citenamefont {Miranda},\ and\ \citenamefont
  {Roig}}]{Masjuan:2023qsp}%
  \BibitemOpen
  \bibfield  {author} {\bibinfo {author} {\bibfnamefont {P.}~\bibnamefont
  {Masjuan}}, \bibinfo {author} {\bibfnamefont {A.}~\bibnamefont {Miranda}},\
  and\ \bibinfo {author} {\bibfnamefont {P.}~\bibnamefont {Roig}},\ }\href
  {https://doi.org/10.1016/j.physletb.2024.138492} {\bibfield  {journal}
  {\bibinfo  {journal} {Phys. Lett. B}\ }\textbf {\bibinfo {volume} {850}},\
  \bibinfo {pages} {138492} (\bibinfo {year} {2024})},\ \Eprint
  {https://arxiv.org/abs/2305.20005} {arXiv:2305.20005 [hep-ph]} \BibitemShut
  {NoStop}%
\bibitem [{\citenamefont {Parker}\ \emph {et~al.}(2018)\citenamefont {Parker},
  \citenamefont {Yu}, \citenamefont {Zhong}, \citenamefont {Estey},\ and\
  \citenamefont {M\"uller}}]{Parker:2018vye}%
  \BibitemOpen
  \bibfield  {author} {\bibinfo {author} {\bibfnamefont {R.~H.}\ \bibnamefont
  {Parker}}, \bibinfo {author} {\bibfnamefont {C.}~\bibnamefont {Yu}}, \bibinfo
  {author} {\bibfnamefont {W.}~\bibnamefont {Zhong}}, \bibinfo {author}
  {\bibfnamefont {B.}~\bibnamefont {Estey}},\ and\ \bibinfo {author}
  {\bibfnamefont {H.}~\bibnamefont {M\"uller}},\ }\href
  {https://doi.org/10.1126/science.aap7706} {\bibfield  {journal} {\bibinfo
  {journal} {Science}\ }\textbf {\bibinfo {volume} {360}},\ \bibinfo {pages}
  {191} (\bibinfo {year} {2018})},\ \Eprint {https://arxiv.org/abs/1812.04130}
  {arXiv:1812.04130 [physics.atom-ph]} \BibitemShut {NoStop}%
\bibitem [{\citenamefont {Morel}\ \emph {et~al.}(2020)\citenamefont {Morel},
  \citenamefont {Yao}, \citenamefont {Clad\'e},\ and\ \citenamefont
  {Guellati-Kh\'elifa}}]{Morel:2020dww}%
  \BibitemOpen
  \bibfield  {author} {\bibinfo {author} {\bibfnamefont {L.}~\bibnamefont
  {Morel}}, \bibinfo {author} {\bibfnamefont {Z.}~\bibnamefont {Yao}}, \bibinfo
  {author} {\bibfnamefont {P.}~\bibnamefont {Clad\'e}},\ and\ \bibinfo {author}
  {\bibfnamefont {S.}~\bibnamefont {Guellati-Kh\'elifa}},\ }\href
  {https://doi.org/10.1038/s41586-020-2964-7} {\bibfield  {journal} {\bibinfo
  {journal} {Nature}\ }\textbf {\bibinfo {volume} {588}},\ \bibinfo {pages}
  {61} (\bibinfo {year} {2020})}\BibitemShut {NoStop}%
\bibitem [{\citenamefont {Giudice}\ \emph {et~al.}(2012)\citenamefont
  {Giudice}, \citenamefont {Paradisi},\ and\ \citenamefont
  {Passera}}]{Giudice:2012ms}%
  \BibitemOpen
  \bibfield  {author} {\bibinfo {author} {\bibfnamefont {G.~F.}\ \bibnamefont
  {Giudice}}, \bibinfo {author} {\bibfnamefont {P.}~\bibnamefont {Paradisi}},\
  and\ \bibinfo {author} {\bibfnamefont {M.}~\bibnamefont {Passera}},\ }\href
  {https://doi.org/10.1007/JHEP11(2012)113} {\bibfield  {journal} {\bibinfo
  {journal} {JHEP}\ }\textbf {\bibinfo {volume} {11}},\ \bibinfo {pages}
  {113}},\ \Eprint {https://arxiv.org/abs/1208.6583} {arXiv:1208.6583 [hep-ph]}
  \BibitemShut {NoStop}%
\bibitem [{\citenamefont {Crivellin}\ \emph {et~al.}(2018)\citenamefont
  {Crivellin}, \citenamefont {Hoferichter},\ and\ \citenamefont
  {Schmidt-Wellenburg}}]{Crivellin:2018qmi}%
  \BibitemOpen
  \bibfield  {author} {\bibinfo {author} {\bibfnamefont {A.}~\bibnamefont
  {Crivellin}}, \bibinfo {author} {\bibfnamefont {M.}~\bibnamefont
  {Hoferichter}},\ and\ \bibinfo {author} {\bibfnamefont {P.}~\bibnamefont
  {Schmidt-Wellenburg}},\ }\href {https://doi.org/10.1103/PhysRevD.98.113002}
  {\bibfield  {journal} {\bibinfo  {journal} {Phys. Rev. D}\ }\textbf {\bibinfo
  {volume} {98}},\ \bibinfo {pages} {113002} (\bibinfo {year} {2018})},\
  \Eprint {https://arxiv.org/abs/1807.11484} {arXiv:1807.11484 [hep-ph]}
  \BibitemShut {NoStop}%
\bibitem [{\citenamefont {Fan}\ \emph {et~al.}(2023)\citenamefont {Fan},
  \citenamefont {Myers}, \citenamefont {Sukra},\ and\ \citenamefont
  {Gabrielse}}]{Fan:2022eto}%
  \BibitemOpen
  \bibfield  {author} {\bibinfo {author} {\bibfnamefont {X.}~\bibnamefont
  {Fan}}, \bibinfo {author} {\bibfnamefont {T.~G.}\ \bibnamefont {Myers}},
  \bibinfo {author} {\bibfnamefont {B.~A.~D.}\ \bibnamefont {Sukra}},\ and\
  \bibinfo {author} {\bibfnamefont {G.}~\bibnamefont {Gabrielse}},\ }\href
  {https://doi.org/10.1103/PhysRevLett.130.071801} {\bibfield  {journal}
  {\bibinfo  {journal} {Phys. Rev. Lett.}\ }\textbf {\bibinfo {volume} {130}},\
  \bibinfo {pages} {071801} (\bibinfo {year} {2023})},\ \Eprint
  {https://arxiv.org/abs/2209.13084} {arXiv:2209.13084 [physics.atom-ph]}
  \BibitemShut {NoStop}%
\bibitem [{\citenamefont {Aoyama}\ \emph {et~al.}(2015)\citenamefont {Aoyama},
  \citenamefont {Hayakawa}, \citenamefont {Kinoshita},\ and\ \citenamefont
  {Nio}}]{Aoyama:2014sxa}%
  \BibitemOpen
  \bibfield  {author} {\bibinfo {author} {\bibfnamefont {T.}~\bibnamefont
  {Aoyama}}, \bibinfo {author} {\bibfnamefont {M.}~\bibnamefont {Hayakawa}},
  \bibinfo {author} {\bibfnamefont {T.}~\bibnamefont {Kinoshita}},\ and\
  \bibinfo {author} {\bibfnamefont {M.}~\bibnamefont {Nio}},\ }\href
  {https://doi.org/10.1103/PhysRevD.91.033006} {\bibfield  {journal} {\bibinfo
  {journal} {Phys. Rev. D}\ }\textbf {\bibinfo {volume} {91}},\ \bibinfo
  {pages} {033006} (\bibinfo {year} {2015})},\ \bibinfo {note} {[Erratum:
  Phys.Rev.D 96, 019901 (2017)]},\ \Eprint {https://arxiv.org/abs/1412.8284}
  {arXiv:1412.8284 [hep-ph]} \BibitemShut {NoStop}%
\bibitem [{\citenamefont {Volkov}(2019)}]{Volkov:2019phy}%
  \BibitemOpen
  \bibfield  {author} {\bibinfo {author} {\bibfnamefont {S.}~\bibnamefont
  {Volkov}},\ }\href {https://doi.org/10.1103/PhysRevD.100.096004} {\bibfield
  {journal} {\bibinfo  {journal} {Phys. Rev. D}\ }\textbf {\bibinfo {volume}
  {100}},\ \bibinfo {pages} {096004} (\bibinfo {year} {2019})},\ \Eprint
  {https://arxiv.org/abs/1909.08015} {arXiv:1909.08015 [hep-ph]} \BibitemShut
  {NoStop}%
\bibitem [{\citenamefont {Volkov}(2024)}]{Volkov:2024yzc}%
  \BibitemOpen
  \bibfield  {author} {\bibinfo {author} {\bibfnamefont {S.}~\bibnamefont
  {Volkov}},\ }\href@noop {} {\  (\bibinfo {year} {2024})},\ \Eprint
  {https://arxiv.org/abs/2404.00649} {arXiv:2404.00649 [hep-ph]} \BibitemShut
  {NoStop}%
\bibitem [{dis()}]{discrepancyres}%
  \BibitemOpen
  \href@noop {} {}\bibinfo {howpublished}
  {\url{https://conference-indico.kek.jp/event/257/contributions/5806/attachments/3737/5122/Muong_2QED2024.pdf}}\BibitemShut
  {NoStop}%
\bibitem [{\citenamefont {Bernabeu}\ \emph {et~al.}(2008)\citenamefont
  {Bernabeu}, \citenamefont {Gonzalez-Sprinberg}, \citenamefont
  {Papavassiliou},\ and\ \citenamefont {Vidal}}]{Bernabeu:2007rr}%
  \BibitemOpen
  \bibfield  {author} {\bibinfo {author} {\bibfnamefont {J.}~\bibnamefont
  {Bernabeu}}, \bibinfo {author} {\bibfnamefont {G.~A.}\ \bibnamefont
  {Gonzalez-Sprinberg}}, \bibinfo {author} {\bibfnamefont {J.}~\bibnamefont
  {Papavassiliou}},\ and\ \bibinfo {author} {\bibfnamefont {J.}~\bibnamefont
  {Vidal}},\ }\href {https://doi.org/10.1016/j.nuclphysb.2007.09.001}
  {\bibfield  {journal} {\bibinfo  {journal} {Nucl. Phys. B}\ }\textbf
  {\bibinfo {volume} {790}},\ \bibinfo {pages} {160} (\bibinfo {year}
  {2008})},\ \Eprint {https://arxiv.org/abs/0707.2496} {arXiv:0707.2496
  [hep-ph]} \BibitemShut {NoStop}%
\bibitem [{\citenamefont {Bernabeu}\ \emph {et~al.}(2009)\citenamefont
  {Bernabeu}, \citenamefont {Gonzalez-Sprinberg},\ and\ \citenamefont
  {Vidal}}]{Bernabeu:2008ii}%
  \BibitemOpen
  \bibfield  {author} {\bibinfo {author} {\bibfnamefont {J.}~\bibnamefont
  {Bernabeu}}, \bibinfo {author} {\bibfnamefont {G.~A.}\ \bibnamefont
  {Gonzalez-Sprinberg}},\ and\ \bibinfo {author} {\bibfnamefont
  {J.}~\bibnamefont {Vidal}},\ }\href
  {https://doi.org/10.1088/1126-6708/2009/01/062} {\bibfield  {journal}
  {\bibinfo  {journal} {JHEP}\ }\textbf {\bibinfo {volume} {01}},\ \bibinfo
  {pages} {062}},\ \Eprint {https://arxiv.org/abs/0807.2366} {arXiv:0807.2366
  [hep-ph]} \BibitemShut {NoStop}%
\bibitem [{\citenamefont {Chen}\ and\ \citenamefont {Wu}(2019)}]{Chen:2018cxt}%
  \BibitemOpen
  \bibfield  {author} {\bibinfo {author} {\bibfnamefont {X.}~\bibnamefont
  {Chen}}\ and\ \bibinfo {author} {\bibfnamefont {Y.}~\bibnamefont {Wu}},\
  }\href {https://doi.org/10.1007/JHEP10(2019)089} {\bibfield  {journal}
  {\bibinfo  {journal} {JHEP}\ }\textbf {\bibinfo {volume} {10}},\ \bibinfo
  {pages} {089}},\ \Eprint {https://arxiv.org/abs/1803.00501} {arXiv:1803.00501
  [hep-ph]} \BibitemShut {NoStop}%
\bibitem [{\citenamefont {Crivellin}\ \emph {et~al.}(2022)\citenamefont
  {Crivellin}, \citenamefont {Hoferichter},\ and\ \citenamefont
  {Roney}}]{Crivellin:2021spu}%
  \BibitemOpen
  \bibfield  {author} {\bibinfo {author} {\bibfnamefont {A.}~\bibnamefont
  {Crivellin}}, \bibinfo {author} {\bibfnamefont {M.}~\bibnamefont
  {Hoferichter}},\ and\ \bibinfo {author} {\bibfnamefont {J.~M.}\ \bibnamefont
  {Roney}},\ }\href {https://doi.org/10.1103/PhysRevD.106.093007} {\bibfield
  {journal} {\bibinfo  {journal} {Phys. Rev. D}\ }\textbf {\bibinfo {volume}
  {106}},\ \bibinfo {pages} {093007} (\bibinfo {year} {2022})},\ \Eprint
  {https://arxiv.org/abs/2111.10378} {arXiv:2111.10378 [hep-ph]} \BibitemShut
  {NoStop}%
\bibitem [{\citenamefont {Janot}(2016)}]{Janot:2015gjr}%
  \BibitemOpen
  \bibfield  {author} {\bibinfo {author} {\bibfnamefont {P.}~\bibnamefont
  {Janot}},\ }\href {https://doi.org/10.1007/JHEP02(2016)053} {\bibfield
  {journal} {\bibinfo  {journal} {JHEP}\ }\textbf {\bibinfo {volume} {02}},\
  \bibinfo {pages} {053}},\ \bibinfo {note} {[Erratum: JHEP 11, 164 (2017)]},\
  \Eprint {https://arxiv.org/abs/1512.05544} {arXiv:1512.05544 [hep-ph]}
  \BibitemShut {NoStop}%
\bibitem [{\citenamefont {Blondel}\ and\ \citenamefont
  {Janot}(2022)}]{Blondel:2021ema}%
  \BibitemOpen
  \bibfield  {author} {\bibinfo {author} {\bibfnamefont {A.}~\bibnamefont
  {Blondel}}\ and\ \bibinfo {author} {\bibfnamefont {P.}~\bibnamefont
  {Janot}},\ }\href {https://doi.org/10.1140/epjp/s13360-021-02154-9}
  {\bibfield  {journal} {\bibinfo  {journal} {Eur. Phys. J. Plus}\ }\textbf
  {\bibinfo {volume} {137}},\ \bibinfo {pages} {92} (\bibinfo {year} {2022})},\
  \Eprint {https://arxiv.org/abs/2106.13885} {arXiv:2106.13885 [hep-ex]}
  \BibitemShut {NoStop}%
\bibitem [{\citenamefont {Tanabashi}\ \emph {et~al.}(2018)\citenamefont
  {Tanabashi} \emph {et~al.}}]{ParticleDataGroup:2018ovx}%
  \BibitemOpen
  \bibfield  {author} {\bibinfo {author} {\bibfnamefont {M.}~\bibnamefont
  {Tanabashi}} \emph {et~al.} (\bibinfo {collaboration} {Particle Data
  Group}),\ }\href {https://doi.org/10.1103/PhysRevD.98.030001} {\bibfield
  {journal} {\bibinfo  {journal} {Phys. Rev. D}\ }\textbf {\bibinfo {volume}
  {98}},\ \bibinfo {pages} {030001} (\bibinfo {year} {2018})}\BibitemShut
  {NoStop}%
\bibitem [{\citenamefont {Gambino}\ and\ \citenamefont
  {Sirlin}(1994)}]{Gambino:1993dd}%
  \BibitemOpen
  \bibfield  {author} {\bibinfo {author} {\bibfnamefont {P.}~\bibnamefont
  {Gambino}}\ and\ \bibinfo {author} {\bibfnamefont {A.}~\bibnamefont
  {Sirlin}},\ }\href {https://doi.org/10.1103/PhysRevD.49.R1160} {\bibfield
  {journal} {\bibinfo  {journal} {Phys. Rev. D}\ }\textbf {\bibinfo {volume}
  {49}},\ \bibinfo {pages} {1160} (\bibinfo {year} {1994})},\ \Eprint
  {https://arxiv.org/abs/hep-ph/9309326} {arXiv:hep-ph/9309326} \BibitemShut
  {NoStop}%
\bibitem [{\citenamefont {Erler}\ and\ \citenamefont
  {Ferro-Hern\'andez}(2018)}]{Erler:2017knj}%
  \BibitemOpen
  \bibfield  {author} {\bibinfo {author} {\bibfnamefont {J.}~\bibnamefont
  {Erler}}\ and\ \bibinfo {author} {\bibfnamefont {R.}~\bibnamefont
  {Ferro-Hern\'andez}},\ }\href {https://doi.org/10.1007/JHEP03(2018)196}
  {\bibfield  {journal} {\bibinfo  {journal} {JHEP}\ }\textbf {\bibinfo
  {volume} {03}},\ \bibinfo {pages} {196}},\ \Eprint
  {https://arxiv.org/abs/1712.09146} {arXiv:1712.09146 [hep-ph]} \BibitemShut
  {NoStop}%
\bibitem [{\citenamefont {Erler}\ \emph {et~al.}(2024)\citenamefont {Erler},
  \citenamefont {Ferro-Hernandez},\ and\ \citenamefont
  {Kuberski}}]{Erler:2024lds}%
  \BibitemOpen
  \bibfield  {author} {\bibinfo {author} {\bibfnamefont {J.}~\bibnamefont
  {Erler}}, \bibinfo {author} {\bibfnamefont {R.}~\bibnamefont
  {Ferro-Hernandez}},\ and\ \bibinfo {author} {\bibfnamefont {S.}~\bibnamefont
  {Kuberski}},\ }\href@noop {} {\  (\bibinfo {year} {2024})},\ \Eprint
  {https://arxiv.org/abs/2406.16691} {arXiv:2406.16691 [hep-ph]} \BibitemShut
  {NoStop}%
\bibitem [{\citenamefont {Becker}\ \emph {et~al.}(2018)\citenamefont {Becker}
  \emph {et~al.}}]{Becker:2018ggl}%
  \BibitemOpen
  \bibfield  {author} {\bibinfo {author} {\bibfnamefont {D.}~\bibnamefont
  {Becker}} \emph {et~al.},\ }\href
  {https://doi.org/10.1140/epja/i2018-12611-6} {\bibfield  {journal} {\bibinfo
  {journal} {Eur. Phys. J. A}\ }\textbf {\bibinfo {volume} {54}},\ \bibinfo
  {pages} {208} (\bibinfo {year} {2018})},\ \Eprint
  {https://arxiv.org/abs/1802.04759} {arXiv:1802.04759 [nucl-ex]} \BibitemShut
  {NoStop}%
\bibitem [{\citenamefont {Benesch}\ \emph {et~al.}(2014)\citenamefont {Benesch}
  \emph {et~al.}}]{MOLLER:2014iki}%
  \BibitemOpen
  \bibfield  {author} {\bibinfo {author} {\bibfnamefont {J.}~\bibnamefont
  {Benesch}} \emph {et~al.} (\bibinfo {collaboration} {MOLLER}),\ }\href@noop
  {} {\  (\bibinfo {year} {2014})},\ \Eprint {https://arxiv.org/abs/1411.4088}
  {arXiv:1411.4088 [nucl-ex]} \BibitemShut {NoStop}%
\bibitem [{\citenamefont {Tiesinga}\ \emph {et~al.}(2021)\citenamefont
  {Tiesinga}, \citenamefont {Mohr}, \citenamefont {Newell},\ and\ \citenamefont
  {Taylor}}]{Tiesinga:2021myr}%
  \BibitemOpen
  \bibfield  {author} {\bibinfo {author} {\bibfnamefont {E.}~\bibnamefont
  {Tiesinga}}, \bibinfo {author} {\bibfnamefont {P.~J.}\ \bibnamefont {Mohr}},
  \bibinfo {author} {\bibfnamefont {D.~B.}\ \bibnamefont {Newell}},\ and\
  \bibinfo {author} {\bibfnamefont {B.~N.}\ \bibnamefont {Taylor}},\ }\href
  {https://doi.org/10.1103/RevModPhys.93.025010} {\bibfield  {journal}
  {\bibinfo  {journal} {Rev. Mod. Phys.}\ }\textbf {\bibinfo {volume} {93}},\
  \bibinfo {pages} {025010} (\bibinfo {year} {2021})}\BibitemShut {NoStop}%
\bibitem [{\citenamefont {Eides}(2019)}]{Eides:2018rph}%
  \BibitemOpen
  \bibfield  {author} {\bibinfo {author} {\bibfnamefont {M.~I.}\ \bibnamefont
  {Eides}},\ }\href {https://doi.org/10.1016/j.physletb.2019.06.011} {\bibfield
   {journal} {\bibinfo  {journal} {Phys. Lett. B}\ }\textbf {\bibinfo {volume}
  {795}},\ \bibinfo {pages} {113} (\bibinfo {year} {2019})},\ \Eprint
  {https://arxiv.org/abs/1812.10881} {arXiv:1812.10881 [hep-ph]} \BibitemShut
  {NoStop}%
\bibitem [{\citenamefont {Czarnecki}\ \emph {et~al.}(2002)\citenamefont
  {Czarnecki}, \citenamefont {Eidelman},\ and\ \citenamefont
  {Karshenboim}}]{Czarnecki:2001yx}%
  \BibitemOpen
  \bibfield  {author} {\bibinfo {author} {\bibfnamefont {A.}~\bibnamefont
  {Czarnecki}}, \bibinfo {author} {\bibfnamefont {S.~I.}\ \bibnamefont
  {Eidelman}},\ and\ \bibinfo {author} {\bibfnamefont {S.~G.}\ \bibnamefont
  {Karshenboim}},\ }\href {https://doi.org/10.1103/PhysRevD.65.053004}
  {\bibfield  {journal} {\bibinfo  {journal} {Phys. Rev. D}\ }\textbf {\bibinfo
  {volume} {65}},\ \bibinfo {pages} {053004} (\bibinfo {year} {2002})},\
  \Eprint {https://arxiv.org/abs/hep-ph/0107327} {arXiv:hep-ph/0107327}
  \BibitemShut {NoStop}%
\bibitem [{\citenamefont {Karshenboim}\ and\ \citenamefont
  {Shelyuto}(2001)}]{Karshenboim:2001yy}%
  \BibitemOpen
  \bibfield  {author} {\bibinfo {author} {\bibfnamefont {S.~G.}\ \bibnamefont
  {Karshenboim}}\ and\ \bibinfo {author} {\bibfnamefont {V.~A.}\ \bibnamefont
  {Shelyuto}},\ }\href {https://doi.org/10.1016/S0370-2693(01)00994-7}
  {\bibfield  {journal} {\bibinfo  {journal} {Phys. Lett. B}\ }\textbf
  {\bibinfo {volume} {517}},\ \bibinfo {pages} {32} (\bibinfo {year} {2001})},\
  \Eprint {https://arxiv.org/abs/hep-ph/0107328} {arXiv:hep-ph/0107328}
  \BibitemShut {NoStop}%
\bibitem [{\citenamefont {Delaunay}\ \emph {et~al.}(2021)\citenamefont
  {Delaunay}, \citenamefont {Ohayon},\ and\ \citenamefont
  {Soreq}}]{Delaunay:2021uph}%
  \BibitemOpen
  \bibfield  {author} {\bibinfo {author} {\bibfnamefont {C.}~\bibnamefont
  {Delaunay}}, \bibinfo {author} {\bibfnamefont {B.}~\bibnamefont {Ohayon}},\
  and\ \bibinfo {author} {\bibfnamefont {Y.}~\bibnamefont {Soreq}},\ }\href
  {https://doi.org/10.1103/PhysRevLett.127.251801} {\bibfield  {journal}
  {\bibinfo  {journal} {Phys. Rev. Lett.}\ }\textbf {\bibinfo {volume} {127}},\
  \bibinfo {pages} {251801} (\bibinfo {year} {2021})},\ \Eprint
  {https://arxiv.org/abs/2106.11998} {arXiv:2106.11998 [hep-ph]} \BibitemShut
  {NoStop}%
\bibitem [{\citenamefont {Mariam}\ \emph {et~al.}(1982)\citenamefont {Mariam}
  \emph {et~al.}}]{Mariam:1982bq}%
  \BibitemOpen
  \bibfield  {author} {\bibinfo {author} {\bibfnamefont {F.~G.}\ \bibnamefont
  {Mariam}} \emph {et~al.},\ }\href
  {https://doi.org/10.1103/PhysRevLett.49.993} {\bibfield  {journal} {\bibinfo
  {journal} {Phys. Rev. Lett.}\ }\textbf {\bibinfo {volume} {49}},\ \bibinfo
  {pages} {993} (\bibinfo {year} {1982})}\BibitemShut {NoStop}%
\bibitem [{\citenamefont {Liu}\ \emph {et~al.}(1999)\citenamefont {Liu} \emph
  {et~al.}}]{Liu:1999iz}%
  \BibitemOpen
  \bibfield  {author} {\bibinfo {author} {\bibfnamefont {W.}~\bibnamefont
  {Liu}} \emph {et~al.},\ }\href {https://doi.org/10.1103/PhysRevLett.82.711}
  {\bibfield  {journal} {\bibinfo  {journal} {Phys. Rev. Lett.}\ }\textbf
  {\bibinfo {volume} {82}},\ \bibinfo {pages} {711} (\bibinfo {year}
  {1999})}\BibitemShut {NoStop}%
\bibitem [{\citenamefont {Strasser}\ \emph {et~al.}(2016)\citenamefont
  {Strasser} \emph {et~al.}}]{Strasser:2016smg}%
  \BibitemOpen
  \bibfield  {author} {\bibinfo {author} {\bibfnamefont {P.}~\bibnamefont
  {Strasser}} \emph {et~al.},\ }\href
  {https://doi.org/10.1007/s10751-016-1331-4} {\bibfield  {journal} {\bibinfo
  {journal} {Hyperfine Interact.}\ }\textbf {\bibinfo {volume} {237}},\
  \bibinfo {pages} {124} (\bibinfo {year} {2016})}\BibitemShut {NoStop}%
\bibitem [{\citenamefont {Friar}\ \emph {et~al.}(1999)\citenamefont {Friar},
  \citenamefont {Martorell},\ and\ \citenamefont {Sprung}}]{Friar:1998wu}%
  \BibitemOpen
  \bibfield  {author} {\bibinfo {author} {\bibfnamefont {J.~L.}\ \bibnamefont
  {Friar}}, \bibinfo {author} {\bibfnamefont {J.}~\bibnamefont {Martorell}},\
  and\ \bibinfo {author} {\bibfnamefont {D.~W.~L.}\ \bibnamefont {Sprung}},\
  }\href {https://doi.org/10.1103/PhysRevA.59.4061} {\bibfield  {journal}
  {\bibinfo  {journal} {Phys. Rev. A}\ }\textbf {\bibinfo {volume} {59}},\
  \bibinfo {pages} {4061} (\bibinfo {year} {1999})},\ \Eprint
  {https://arxiv.org/abs/nucl-th/9812053} {arXiv:nucl-th/9812053} \BibitemShut
  {NoStop}%
\bibitem [{\citenamefont {Meyer}\ \emph {et~al.}(2000)\citenamefont {Meyer}
  \emph {et~al.}}]{Meyer:1999cx}%
  \BibitemOpen
  \bibfield  {author} {\bibinfo {author} {\bibfnamefont {V.}~\bibnamefont
  {Meyer}} \emph {et~al.},\ }\href
  {https://doi.org/10.1103/PhysRevLett.84.1136} {\bibfield  {journal} {\bibinfo
   {journal} {Phys. Rev. Lett.}\ }\textbf {\bibinfo {volume} {84}},\ \bibinfo
  {pages} {1136} (\bibinfo {year} {2000})},\ \Eprint
  {https://arxiv.org/abs/hep-ex/9907013} {arXiv:hep-ex/9907013} \BibitemShut
  {NoStop}%
\bibitem [{\citenamefont {Crivelli}(2018)}]{Crivelli:2018vfe}%
  \BibitemOpen
  \bibfield  {author} {\bibinfo {author} {\bibfnamefont {P.}~\bibnamefont
  {Crivelli}},\ }\href {https://doi.org/10.1007/s10751-018-1525-z} {\bibfield
  {journal} {\bibinfo  {journal} {Hyperfine Interact.}\ }\textbf {\bibinfo
  {volume} {239}},\ \bibinfo {pages} {49} (\bibinfo {year} {2018})},\ \Eprint
  {https://arxiv.org/abs/1811.00310} {arXiv:1811.00310 [physics.atom-ph]}
  \BibitemShut {NoStop}%
\bibitem [{\citenamefont {Eides}\ and\ \citenamefont
  {Shelyuto}(2016)}]{Eides:2016qhf}%
  \BibitemOpen
  \bibfield  {author} {\bibinfo {author} {\bibfnamefont {M.~I.}\ \bibnamefont
  {Eides}}\ and\ \bibinfo {author} {\bibfnamefont {V.~A.}\ \bibnamefont
  {Shelyuto}},\ }\href {https://doi.org/10.1142/S0217751X16450342} {\bibfield
  {journal} {\bibinfo  {journal} {Int. J. Mod. Phys. A}\ }\textbf {\bibinfo
  {volume} {31}},\ \bibinfo {pages} {1645034} (\bibinfo {year}
  {2016})}\BibitemShut {NoStop}%
\end{thebibliography}%

\end{document}